\documentclass[rmp, aps, preprint, longbibliography]{revtex4-1}
\usepackage{epsfig}
\usepackage{amssymb}
\usepackage{epstopdf}
\usepackage{amsmath}
\usepackage{bm}

\usepackage{xcolor}
\usepackage{soul}

\newcommand{\eq}[1]{{Eqn.~(\ref{#1})}}

\begin{document}

\count\footins = 1000

\title{{\it Colloquium}: Bell's Theorem and Locally-Mediated Reformulations of Quantum Mechanics}
\author{K.B. Wharton}
\email{kenneth.wharton@sjsu.edu}
\affiliation{Department of Physics and Astronomy, San Jos\'{e} State University, San Jos\'{e}, CA 95192-0106}
\author{N. Argaman}
\email{argaman@mailaps.org}
\affiliation{Department of Physics, Nuclear Research Center---Negev, P.O. Box 9001, Be'er Sheva 84190, Israel}
\date{\today}       

\begin{abstract}

Bell's Theorem rules out many potential reformulations of quantum mechanics, but within a generalized framework, it does not exclude all ``locally-mediated'' models.  Such models describe the correlations between entangled particles as mediated by intermediate parameters which track the particle world-lines and respect Lorentz covariance.  These locally-mediated models require the relaxation of an arrow-of-time assumption which is typically taken for granted.  Specifically, some of the mediating parameters in these models must functionally depend on measurement settings in their future, \emph{i.e.}, on input parameters associated with later times.  This option (often called ``retrocausal'') has been repeatedly pointed out in the literature, but the exploration of explicit locally-mediated toy-models capable of describing specific entanglement phenomena has begun only in the past decade.  A brief survey of such models is included here.  These models provide a continuous and consistent description of events associated with spacetime locations, with aspects that are solved ``all-at-once'' rather than unfolding from the past to the future.  The tension between quantum mechanics and relativity which is usually associated with Bell's Theorem does not occur here.  Unlike conventional quantum models, the number of parameters needed to specify the state of a system does not grow exponentially with the number of entangled particles.  The promise of generalizing such models to account for all quantum phenomena is identified as a grand challenge.

\end{abstract}


\maketitle

\tableofcontents

\newpage
\hyphenpenalty=1400

\section{Introduction}

Bell's Theorem places a strong restriction on reformulations of Quantum Mechanics (QM): any mathematical model which produces the same output predictions as QM, given the same inputs, must violate {\bf Local Causality}%
\footnote{Boldface is used for mathematical conditions explicitly defined in the following sections.} \cite{bell1964}.
In this sense, QM is ``nonlocal,'' but locality is not a simple yes/no question; QM is still ``local'' according to the \emph{operational} definition, as it does not allow signaling at a distance, or outside the future lightcone.  Thus, QM may even be compatible with a generalized non-operational definition of ``locality,'' not as strict as {\bf Local Causality}, but in spirit with Einstein's arguments against action-at-a-distance.  This Colloquium will examine a category of potential reformulations of QM which are as ``local'' as allowed by Bell's Theorem.

In order to assess a model's ``locality,'' even at the operational level of inputs and outputs, the model must define its spacetime-based parameters [those which \textcite{bell1976b} called ``local beables''].  Such parameters are mathematical variables which are clearly associated with a specific time and place, such as the local values of classical fields, $Q(\bm{x},t)$.  This association allows concepts of ``locality'' to be meaningfully applied.  

Conventional QM utilizes inputs corresponding to values that can be controlled by experimental physicists, including the settings of preparation and measurement devices, and it predicts the probability distribution of the values of measureable outputs.  Reformulations of QM must be operationally equivalent, utilizing these same inputs and outputs, and providing the same predictions.  At minimum, this requires the use of spacetime-based parameters for the inputs and the outputs.

These input and output parameters do not continuously span the intermediate spacetime regions where preparations and measurements are not performed; non-operational definitions of ``locality'' (such as {\bf Local Causality}) concern parameters associated with these intermediate regions.  If a model has no spacetime-based parameters associated with these regions, it will be nonlocal according to these definitions, \emph{i.e.}, it will have unmediated action-at-a-distance.

Many physicists see Bell's Theorem as a reason {\em not} to introduce such mediating parameters [see, \emph{e.g.}, \textcite{mermin1986}].  It is difficult to map an entangled configuration-space wavefunction $\psi(\bm{x}_1,\bm{x}_2,t)$ onto spacetime-based parameters $Q(\bm{x},t)$ \cite{norsen2010,stoica2019}, and even if there were such a mapping,
Bell's Theorem tells us that no reformulation could possibly conform to {\bf Local Causality}.  However, such a viewpoint presumes that there is no other type of ``locality'' worth saving, no subset of assumptions inside of {\bf Local Causality} which might be beneficial for some future reformulation of QM.  

In fact, there exists a category of quantum reformulations for which an essential aspect of ``locality'' can be retained [see, \emph{e.g.}, \textcite{costa1953,price1997,argaman2010}].  These models utilize spacetime-based parameters associated with intermediate regions between preparations and measurements, allowing these models to be ``locally mediated'', in the sense that correlations cannot be introduced or altered except via intermediate spacetime-based mediators.  This condition will be explicitly defined in the next section, using the term {\bf Continuous Action} to contrast with the phrase ``action-at-a-distance''.  We are most interested in cases where this local mediation is always restricted to time-like or light-like worldlines, allowing those models to also respect Lorentz covariance.  

Such locally-mediated reformulations of QM must violate a certain time-asymmetric assumption inherent to {\bf Local Causality}.  Specifically, the relevant assumption presumes that no model parameter associated with time $t$ can be dependent upon model inputs associated with times greater than $t$.  The most ``local'' reformulations of QM---those with {\bf Continuous Action}---violate this assumption, and are therefore future-input dependent.  While some may view such ``retrocausality'' as unreasonable, it is emphasized that only models with the same predictions as QM are of interest here, with no signaling into the past [see, \emph{e.g.}, \textcite{price1997}].  

If one considers that inputs to a model also include boundary conditions, it is evident that future-input dependent models are ubiquitous throughout physics.  For example, models employing the stationary action principle fall in this category -- mathematical inputs constrain both initial and \emph{final} parameters, and the model determines the classical history at intermediate times.  Any calculation of a closed-timelike-curve in general relativity requires a similar all-at-once analysis.  Quantum future-input dependent models, such as the Transactional Interpretation \cite{cramer1980} and the Two-State-Vector Formalism \cite{aharonov1991}, have also been developed, motivated primarily by time-symmetry rather than locality.

Attempts to develop a locally-mediated account of quantum entanglement using future-input dependence have been promoted by a number of forward-thinking authors, beginning even before Bell's work \cite{costa1953,costa1977,costa1979,pegg1982,sutherland1983,price1984}.  Still, mathematical future-input dependent models reproducing the QM predictions for entangled particles, while explicitly maintaining local mediation, have been put forward mainly in the last decade.  One purpose of this Colloquium is to survey the admittedly modest achievements of this recent and still-developing line of enquiry, and to indicate some intriguing directions for further exploration.

A formal development of these arguments will also result in a useful categorization of all the ways in which a reformulation of QM can violate {\bf Local Causality} (we say that such models are ``Bell-compatible'').  This is accomplished by presenting the assumptions of Bell's Theorem in terms consistent with the recently developed framework of ``causal models'' \cite{pearl2009}, which emphasizes the role of QM's controllable inputs.  Bell himself spoke of the special importance of inputs, calling them ``free external variables in addition to those internal to and conditioned by the theory'' \cite{bell1977}.  Unfortunately, the mathematics of causal models was not well-developed during his lifetime, and \textcite{bell1976b,bell1981,bell1990} adopted a neutral notation, \emph{e.g.}, $\{ A | a, \lambda \}$ for the probability distribution of an output $A$ given an input $a$ and an internal parameter $\lambda$.  Instead, we will denote this by $p_a(A|\lambda)$, emphasizing the input status of $a$, and allowing a clear categorization of Bell-compatible reformulations of QM (while setting aside extraneous issues such as ``superdeterminism'').

Developing reformulations of existing theories has historically been very useful -- think of the advances of Lagrangian and Hamiltonian classical mechanics.  In quantum theory, the path integral has similarly led to new insights, and there is no indication that this strategy of seeking further reformulations has run its course \cite{feynman1964}.  In particular, an alternative quantum model with parameters restored to functions on spacetime, instead of a multi-dimensional configuration space (or a Hilbert space), would have significant advantages.  Such a model would have a natural interpretation, with one allowed combination of the spacetime-based parameters corresponding to physical reality, and all other combinations being mere possibilities (as in classical statistical mechanics).  As a result, the number of parameters describing an actual system would grow linearly (rather than exponentially) with the extent of that system.  This would substantially lessen the disconnect between quantum theory and our linearly-scaling classical theories of relativistic spacetime.

Note that a successful reformulation of QM in terms of spacetime-based parameters would certainly not imply that quantum theory was incorrect.  Quantum states could still represent our best possible knowledge about measurable aspects of those parameters, given accessible information.  In this case, quantum states could be viewed as states of knowledge, a popular perspective in the field of quantum information \cite{caves2002,spekkens2007,leifer2013}.

The next section carefully walks through Bell's Theorem, identifying all the assumptions leading to the contradiction with quantum phenomena.  Section III then categorizes Bell-compatible reformulations of QM.  Several examples of locally-mediated toy models are detailed in Section IV; those who would like to look at a concrete mathematical model, rather than follow general reasoning, are referred to the model of Section IV.B, for which a detailed derivation is given in the appendix.  Section V discusses the approach and indicates avenues for further development.  Alternative approaches are briefly discussed in Section VI\@.  Section VII provides the conclusion, encouraging future development of locally-mediated reformulations of QM.

\section{Bell's Theorem}

Our first task is to prove Bell's Theorem.  Starting with a certain set of natural assumptions, we will give a mathematical proof of a Bell inequality -- specifically, the Clauser-Horne-Shimony-Holt (CHSH) inequality \cite{clauser1969}.  This inequality can be tested operationally (without reference to any underlying theory), and it is experimentally violated, just as predicted by QM.  It follows that for any model of these phenomena, at least one of the assumptions which lead to Bell's Theorem must be violated.  All such ``Bell-compatible'' models can then be usefully categorized in terms of which assumptions are relaxed.  

The analysis is divided into the following Subsections: the first defines the framework rules for the models to be discussed, the second lists the relevant reasonable-but-optional assumptions that could characterize such models, the third provides some historical context, and the fourth provides a derivation of the Theorem. 

\subsection{Framework: spacetime-based models}
\label{sec:framework}

In all of physics, one uses mathematical models to generate falsifiable predictions which can be compared with empirical observations.  The sort of models that accomplish this are essentially functions that take some parameters as inputs and generate other parameters as outputs.\footnote{We use the term ``parameter'' instead of ``variable'', as the latter sometimes implies a time-dependent quantity, while inputs and outputs are generally localized in time, as well as space.}
We are therefore interested in models which come with well-defined {\bf input parameters} (``inputs'' for short), which will be denoted by the set $I$, and also well-defined {\bf output parameters} (``outputs''), denoted by $O$.  Models can have other parameters in addition to the inputs and outputs, and the set of these will be denoted by $U$.  We will often discuss the set of all non-input parameters $Q$ (the union of $O$ and $U$).  Parameters here are not limited to simple scalars---vectors, or more complicated mathematical constructs such as functions may be utilized.

As discussed in the Introduction, we are interested in models of spacetime-based parameters, each associated with a particular location in ordinary spacetime.  Examples of such parameters include the values of physical fields, such as $\bm{E}(\bm{x},t)$ in classical electromagnetism and $g^{\mu\nu}(x^\gamma)$ in general relativity.  Other examples include instrument settings and measurement results, which are associated with definite regions rather than points in spacetime.  These parameters correspond to what \textcite{bell1976b} called ``local beables'' (pronounced be-ables).  Unless otherwise noted, our use of the term ``parameters'' will be restricted to spacetime-based parameters, including the sets $I$ and $Q$.

Of course, some models employ additional mathematical entities which are not spacetime-based.  For example, for $N>1$, the $N$-particle configuration-space wavefunction in QM is comprised of values that do not correspond to particular locations in spacetime.  For the purposes of Bell's analysis and the below discussion, non-spacetime-based parameters such as configuration-space wavefunctions are simply omitted from $Q$, even if they are mathematically utilized in a given model.  It is also possible to construct non-localized parameters out of spacetime-based parameters, such as the total energy of an extended system, but such values are not to be included as elements of $I$ or $Q$.

Deterministic models are those for which specification of all inputs $I$, including boundary conditions and external forces (if present) always exactly determines the non-input parameters $Q$.  Stochastic models do not predict unique values for $Q$, but for any full set of inputs, the model assigns a probability for every possible combination of non-input parameters.  Thus, a fully-specified mathematical model can always be written as $P_I(Q)$, a unique joint probability distribution function for the set of non-input parameters, given specific values for the inputs.%
\footnote{In many cases, ``distribution functional'' rather than ``function'' should be used here, as $Q$ itself typically includes fields, functions of spacetime.  Similarly, when $Q$ is continuous, $P_I(Q)$ denotes probability densities rather than probabilities.}
For deterministic models, these distributions are $\delta$-functions, but the analysis is not limited to such cases.  This definition, which suffices for the present purposes, is minimal in the sense that it does not include the details regarding how one parameter is deduced from another within the model, nor the physical interpretation of a model.
 
According to the standard rules for probabilities, the full joint probability distribution $P_I(Q)$ of all the non-input parameters of a model can be used to generate marginal distributions, $P_I(Q_1)$, for any subset $Q_1 \subset Q$.  It also generates conditional probabilities, $P_I(Q_1|q_2)$, where $q_2$ are specific values of parameters in another subset, $Q_2$.  In some cases, a model may predict that $Q_1$ and $Q_2$ are statistically independent, meaning that $P_I(Q_1,Q_2) = P_I(Q_1) P_I(Q_2)$.  When statistical independence holds, knowledge of the values of parameters in $Q_2$ does not inform the marginal $P_I(Q_1)$, as represented by the condition $P_{I}(Q_1|Q_2)=P_{I}(Q_1)$.

Two models that use identical input and output sets $I$ and $O$ (when applied to a given system) and that also have the same marginal output probabilities $P_I(O)$ are said to be in {\bf agreement}: they always yield the same joint probability of the output parameters for a given set of inputs.  Note that {\bf agreement} does not preclude different predictions at the level of $P_I(Q)$.  Two models can even be in {\bf agreement} if they utilize different parameters $U$ in $Q$.  For example, in Classical Electromagnetism (CEM), one can change the gauge condition on parameters corresponding to the electromagnetic potentials without changing observable model predictions.  The discussion below treats any such parameter-changing reformulations as different models, because they might generally have different properties at the level of non-observable parameters.%
\footnote{In the CEM example, the use of Coulomb-gauge potentials as parameters will generally not respect the {\bf Local Causality} condition defined below, because those potentials can change instantaneously over all space.}
(As we shall see, even changing the associated spacetime location of a parameter can significantly change the model.)

In the following we will focus on models which are in {\bf agreement} with QM, at least for a specific setup under consideration.  Such models are guaranteed to share the empirical success of QM, but are strictly constrained by Bell's Theorem.

\subsection{Physical assumptions}
\label{sec:assumptions}

The following properties may or may not hold for any specific mathematical model, allowing for a categorization of models into classes and sub-classes.  In order to maintain an appropriate scope, we here define only key properties which play significant roles in the discussion to follow, with a few more given in Section~\ref{sec:more_assumptions}.  For example, the need to formally define relativistic covariance of models does not arise here, although the lightcones of Minkowski spacetime do play a role.

\subsubsection{Continuous Action {\rm ({\bf CA})}}

Instead of beginning with Bell's approach to defining locality, we first define a weaker condition, {\bf Continuous Action} ({\bf CA}), that encodes the spirit of no-action-at-a-distance without requiring any lightcone structure from relativity, or even a distinction between past and future.  As shown in Figure 1(a), consider spacetime regions $\bf{1}$ and $\bf{2}$, with $\bf{1}$ completely surrounded by a screening region $\bf{S}$.  This is not merely a spatial region; $\bf{S}$ spans the past and future of $\bf{1}$ as well as its spatial extent.  We will denote the set of all inputs in regions $\bf{1}$ and $\bf{2}$ by $I_1$ and $I_2$, respectively.  If there are any additional inputs, besides $I_1$ and $I_2$, their values are assumed to be fixed in the definitions below. The non-input parameters in each region are denoted by the corresponding $Q_1$, $Q_2$, $Q_S$.

Loosely speaking, a mathematical model violates {\bf CA} if it has unmediated ``action-at-a-distance'', \emph{i.e.}, if changes in $\bf{2}$ can be associated with changes in $\bf{1}$ without being also associated with changes within $\bf{S}$.  For example, a {\bf CA}-respecting model of a light switch in $\bf{2}$ correlated with a lamp in $\bf{1}$ must include a description of the mediating parameters (\emph{e.g.}, the currents in the wires) in the intermediate screening region $\bf{S}$. In such a model, knowledge of the values of all the parameters in $\bf{S}$ makes additional information regarding $\bf{2}$ \emph{redundant} for the purpose of predicting what happens in region $\bf{1}$.

\begin{figure}[t]
\centerline{\includegraphics[width=.5\textwidth]{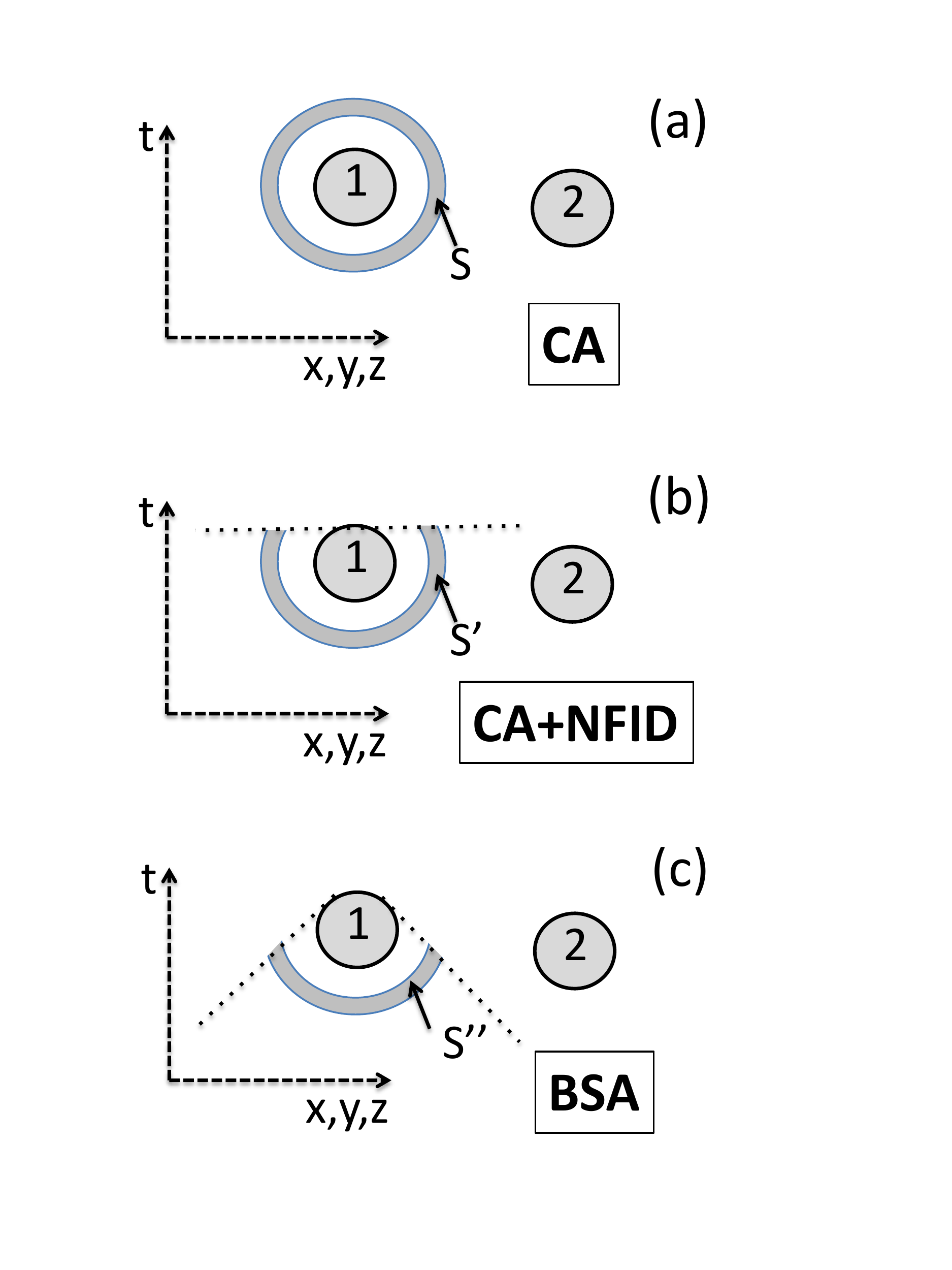}}
\label{Figure:Fig1}
\caption{The screening regions $\bf{S}$ used in different assumptions of ``locality''.  Given all modeled parameters in the screening region, a screened model will assign the same probabilities for parameters in region {\bf 1}, regardless of  additional knowledge of parameters in region {\bf 2}.  Figure 1a shows the most general case of {\bf Continuous Action} ({\bf CA}); Figure 1b breaks time-symmetry by adding the {\bf No Future-Input Dependence} ({\bf NFID}) assumption, and Figure 1c references the lightcones according to {\bf Bell's Screening Assumption} ({\bf BSA}).}
\end{figure}

Mathematically, {\bf CA} corresponds to the condition
\begin{equation}
\label{eq:shield_full}
P_{I_1,I_2}(Q_1|Q_2,Q_S) = P_{I_1}(Q_1|Q_S),
\end{equation}
for all combinations of the parameters in the regions depicted in Figure 1(a).  This equation says that $P_{I_1}(Q_1|Q_S)$ is both {\em statistically independent} of $Q_2$, and {\em functionally independent} of $I_2$.  When this occurs, we say that $\bf{S}$ ``screens'' $\bf{1}$ from $\bf{2}$.\footnote{The word ``shields'' is often used in the literature, including \textcite{bell1990}, instead of ``screens.''}  For {\bf CA} models this equality is required to hold for all simply connected, non-overlapping regions $\bf{1}$, $\bf{2}$, and $\bf{S}$, for which $\bf{S}$ completely separates $\bf{1}$ from $\bf{2}$ and is nowhere vanishingly thin.  As there is no essential difference between regions $\bf{1}$ and $\bf{2}$, a model with {\bf CA} also must have $\bf{S}$ screen $\bf{2}$ from $\bf{1}$.

Readers familiar with probabilistic modeling will notice that the role of the screening region $\bf{S}$ in {\bf CA} is analogous to that of a ``Markov blanket,'' a term coined by \textcite{pearl1988}.   We avoid using this terminology not only because of required minor adjustments (discretizing the model in spacetime onto a set of nodes; properly representing the role of inputs), but primarily because many physicists might be misled---Markov's name would likely be immediately associated with Markov processes, which propagate step by step from the past to the future, subject to a particularly strong arrow of time.  This is exactly the opposite of our purpose here---generalizing ``no action at a distance'' to situations in which time symmetry is not broken at all, or is broken in a much weaker manner.  Restricting attention to Markov processes would be an \emph{additional} assumption---limiting attention to directed acyclic graphs with the directions of all edges determined by temporal order---one that will now be formally defined. 

\subsubsection{No Future-Input Dependence {\rm ({\bf NFID})}}

There is a well-known tension between the time-symmetric equations characteristic of fundamental physical theories and the time-asymmetric manner in which models are utilized.  For example, if one takes wavefunction ``collapse'' to be physically meaningful, this process defines a preferred direction of time, breaking the time symmetry evident in unitary evolution.  More generally, a preferred direction of time is commonly chosen by limiting attention to models in which all events up to a time $t'$ can be evaluated without regards to events in the future of $t'$: 

\begin{quote}
{\bf No Future-Input Dependence} ({\bf NFID}) holds for a mathematical model $P_I(Q)$ if, for any time $t'$ included in the relevant spacetime region, there exists a restricted model $P'_{I'}(Q')$, where $I'$ is the set of all inputs belonging to times up to $t'$ and $Q'$ is the set of all non-input parameters up to $t'$, such that
\begin{equation}
\label{eq:NFID}
P_I(Q')=P'_{I'}(Q')
\end{equation}
for all possible values of the parameters in $I$ and $Q'$.  
\end{quote}
In other words, {\bf NFID} means the marginal $P_I(Q')$ is functionally independent of future inputs.

When combined with {\bf CA}, the assumption of {\bf NFID} implies that there is no need to consider any parts of the screening region $\bf{S}$ that lie in the future of both regions {\bf 1} and {\bf 2}.  As shown in Figure 1(b), if {\bf CA} holds for $P$ and $P'$ of a model respecting {\bf NFID}, then the smaller region ${\bf S'}$ also screens ${\bf 1}$ from ${\bf 2}$, $P_{I_1,I_2}(Q_1|Q_2,Q_{S'}) = P_{I_1}(Q_1|Q_{S'})$.

\subsubsection{Bell's Screening Assumption {\rm ({\bf BSA})}}

If one accepts both of the above assumptions ({\bf CA} and {\bf NFID}), and is furthermore interested in modeling only screening regions that remain applicable in \emph{all} reference frames, it becomes appropriate to ignore any portion of $\bf{S}$ that is spacelike separated from both\footnote{Note that it is not sufficient to restrict the screening region to lie in the past lightcone of region {\bf 1}; it must completely screen {\bf 1} from the overlap of the past lightcones of {\bf 1} and {\bf 2} [see, \emph{e.g.}, note 7 in \textcite{bell1986}].} regions {\bf 1} and {\bf 2}.  This leads to the smaller region $\bf{S''}$ shown in Figure 1(c).  \textcite{bell1990} proposed that this smaller region, $\bf{S''}$, should screen region {\bf 1} from region {\bf 2}:
\begin{equation}
\label{eq:BL}
P_{I_1,I_2}(Q_1|Q_2,Q_{S''}) = P_{I_1}(Q_1|Q_{S''}).
\end{equation}
It is important to note that this screening condition does not imply that parameters in {\bf 1} are {\em independent} of parameter values in {\bf 2}---merely that the latter values are redundant, given the specification of all model parameters in $\bf{S''}$.  We will call \eq{eq:BL} {\bf Bell's Screening Assumption} ({\bf BSA}).

\subsubsection{Local Causality}

Models which conform to both {\bf BSA} and {\bf NFID} are unable to describe certain quantum phenomena, as Bell's Theorem establishes, and will be proved in Section~II.D below.  We will define this important combination of assumptions as {\bf Local Causality}.%
\footnote{The freedom in choosing the region $\bf{S}$ in the definition of {\bf CA} is reflected in the definitions used for ``Local Causality'' (or ``Einstein Locality,'' or ``Local Realism'').  The present Figure~1 resembles Figure~6.4 of \textcite{bell1990}, the ``screening region'' was effectively the entire past of $\bf 1$ and $\bf 2$ in \textcite{bell1981}, and \textcite{bell1976b} used the overlap of their past lightcones.  This affects the identification of $\lambda$ in the separability condition below, Eqn.~(\ref{eq:Sep}), but the subsequent derivation is unchanged.}
This definition may cause some initial confusion, because {\bf Local Causality} is often identified with \eq{eq:BL} in the literature, which is formally just {\bf BSA}.  However, in essentially all cases in which this is done, the authors are presupposing {\bf NFID}, {either explicitly or implicitly,} and the addition of this assumption turns {\bf BSA} into {\bf Local Causality}.  \textcite{bell1990} himself introduced {\bf BSA} after clearly assuming the past-to-future causal structure associated with {\bf NFID} [see Figure 6.3 there], and used the term {\bf Local Causality} to convey this combination, often using the shorter ``locality'' as a synonym.

The reader should be cautioned about interpreting the phrase ``Local Causality'' as being the simple conjunction of ``locality'' and ``causality''.  There are many different meanings that could be ascribed to both of these words (we have already seen three different notions of locality in Fig.~1 above).   All that is needed in the present analysis is that {\bf Local Causality} means the well-defined assumptions {\bf NFID} and {\bf BSA}.

An important condition which follows from {\bf NFID} (or from {\bf Local Causality}), but not {\bf BSA} alone, can be derived by applying it to the $S''$ region from Figure 1(c).  Requiring the probabilities of parameters to be independent of future inputs, and choosing $S''$ to lie entirely in the past of all of regions $\bf{1}$ and $\bf{2}$ (in some reference frame where {\bf NFID} holds), one obtains the functional independence relation
\begin{equation}
\label{eq:SI2}
P_{I_1,I_2}(Q_{S''}) = P(Q_{S''}) .
\end{equation}
A variant of this condition will play an important role in the proof of Bell's Theorem below.

\subsection{Historical interlude}
\label{sec:interlude}

At this point it is appropriate to emphasize how natural it is to assume that all of the above conditions, summarized by {\bf Local Causality}, should hold in any detailed model describing real physics.  It is convenient to do so by referencing Einstein and Bohr.

At the 1927 Solvay conference, Einstein noted that there were two possible ``conceptions'' of the single-particle quantum wavefunction $\psi(x,t)$ \cite{bacciagaluppi2009}. If viewed as a set of spacetime-based parameters $Q(x,t)$, the requirement that only a single particle is eventually measured implies some form of wavefunction collapse that, for Einstein, ``implies to my mind a contradiction with the postulate of relativity.''  Instead, he advocated a conception where ``one does not describe the process solely by the Schr\"odinger wave," effectively pointing out the possibility that additional hidden parameters could indicate the particle's actual location.

In 1935, Einstein, Podolsky and Rosen \nocite{EPR1935} (EPR) extended this analysis to a two-particle system (of a type to be analyzed below), and reached the logical conclusion that violations of {\bf Local Causality} could \emph{only} be avoided by adding new hidden parameters.  If known, these new parameters would allow one to determine the outcomes in more detail than is possible within QM\@.  EPR concluded that QM gave an incomplete description.

EPR did not use the formal mathematical language of Bell's analysis.  Instead, they implied the existence of spacetime-based parameters $Q$ that encoded ``{\it an element of physical reality}'' (italics in original, here and below), and deduced that hidden $Q$'s must be present in a complete theory, because in some cases it was possible to ``{\it predict with certainty ... the value of a physical quantity},'' such as position or momentum, ``{\it without in any way disturbing a system}.''  

\textcite{bohr1935} responded quickly to EPR, defending the completeness of QM on the basis of the notion of complementarity he had developed earlier (in connection with the quantum uncertainty principle).  He advocated ``a radical revision of our attitude towards the problem of physical reality,'' and argued that the phrase ``without in any way disturbing a system'' used by EPR ``contains an ambiguity.''   

Bohr considered in detail a situation in which the properties of a particle can be discerned by first allowing it to pass through a slit in a diaphragm, and later making a ``{\it free choice}'' of measuring either the momentum or the position of the diaphragm.  (It is remarkable, especially in the context of the present work, that the guarantee for ``without in any way disturbing'' was spatial separation for EPR, but temporal order for Bohr.)  ``Of course there is $\dots$ no question of a mechanical disturbance $\dots$ during the last critical stage of the measuring procedure,'' he wrote.  But as one can only measure either the position or the momentum of the diaphragm, ``even at this stage'' there still might be ``{\it an influence on $\dots$ the possible types of predictions regarding the future behavior of the system.}''  
Bohr thus advocated\footnote{We believe it is appropriate to interpret Bohr in this manner, but acknowledge that it is probably impossible to uncontroversially translate his writing into the formal language introduced later.} accepting some violations of {\bf Local Causality} which are present in the formalism of QM, while at the same time excluding other violations---those corresponding to a ``mechanical disturbance''.  (Similarly, his notion of ``completeness'' clearly differs from that of EPR.)

Most physicists simply adopted Bohr's complementarity, either in its original form or a variant \cite{bell1992}, and continued to develop and apply QM to a variety of physical systems \cite{bellHoP,mermin1986}. But Einstein was not convinced.  Summarizing the situation in 1948, he wrote \cite{born1971}:
\begin{quote}
[T]hose physicists who regard the descriptive methods of quantum mechanics as definitive in principle would $\dots$ drop the requirement $\dots$ for the independent existence of the physical reality present in different parts of space. $\dots$  [W]hen I consider the physical phenomena known to me, and especially those which are being so successfully encompassed by quantum mechanics, I still cannot find any fact anywhere which would make it appear likely that [that] requirement will have to be abandoned.  I am therefore inclined to believe that the description of quantum mechanics $\dots$ has to be regarded as an incomplete and indirect description of reality, to be replaced at some later date by a more complete and direct one.
\end{quote}
Here Einstein is essentially advocating for models to be built from spacetime-based parameters $Q$, while offering the opinion that other physicists had prematurely abandoned this possibility.  But there was indeed a ``fact'' that he was not aware of, a theorem that would be proved by Bell in 1964 (sadly, after both Einstein and Bohr had passed away).  We now turn to Bell's Theorem, and the fact that all models in {\bf agreement} with QM must violate the package of assumptions that is {\bf Local Causality}.  Subsequently, we will address the question of whether the hidden-parameter models advocated by Einstein should still be pursued, even given the necessary violation of {\bf Local Causality}.

\subsection{Statement and proof}
\label{sec:theorem}

Bell's Theorem demonstrates that:
\begin{quote} 
No model conforming with {\bf Local Causality} can be in {\bf agreement} with QM.
\end{quote}
It is to be emphasized that the disagreement is not only with the predictions of QM, but also with the results of empirical observations -- experiments which have been performed.  The proof below is based on the CHSH inequality \cite{clauser1969}, which concerns a particular application of QM to the experimental scenario shown in Figure 2.  Specifically, a source emits a pair of particles, and these are later analyzed and detected in spacelike-separated regions $\bf{1}$ and $\bf{2}$.

\begin{figure}[t]
\centerline{\includegraphics[width=0.4\textwidth]{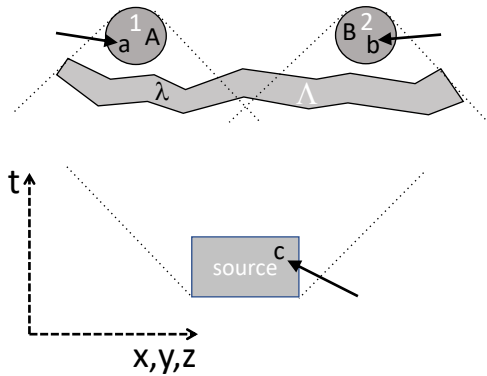}}
\label{Figure:Fig2}
\caption{The essential geometry of a Bell-type experiment.  The parameters $a,b,c$ are inputs; the arrows indicate that their values come from outside the model.  The parameters $A$ and $B$ are observable outputs. $\lambda$ is the set of all localized model parameters in the region $\bm{\Lambda}$, which screens regions $\bm{1}$ and $\bm{2}$ from the overlap of their backward lightcones.}
\end{figure}

Mathematical models describing such situations will have an input parameter $c$ specifying the particular settings/arrangement of the common source of the two particles.  Additional input parameters $a$ and $b$ specify the settings/arrangement of the detectors in region $\bf{1}$ and region $\bf{2}$ respectively.  The results of the experiment are the output parameters $A$ in region $\bf{1}$ and $B$ in region $\bf{2}$.  The set of all the model's spacetime-based parameters in region $\bf{\Lambda}$ is denoted by $\lambda$.  The parameters $a$, $b$ and $c$ are inputs, and $A$ and $B$ are outputs, just as in QM\@.  The set $\lambda$ can be quite general, with possibilities ranging from complex combinations of functions and operators to the simplest possibility: the empty set, for the case with no spacetime-based parameters associated with the region $\bf{\Lambda}$.

The proof of the CHSH inequality, following \textcite{bell1976b,bell1981,bell1990} and \textcite{peres1978}, proceeds in the next two subsections by only assuming {\bf Local Causality}, without making any reference to QM\@.  The third subsection then proves Bell's Theorem by comparing this Bell inequality with quantum theory and experiments. (A disadvantage of Bell's original 1964 proof is discussed in Section~\ref{sec:misconceptions}.) 

\subsubsection{Bell's separability condition}

Any mathematical model capable of producing predictions for the setup of Figure 2 will provide a joint probability distribution $P_{a,b,c}(A,B,\lambda)$.  The marginal distribution $P_{a,b,c}(A,B)$ can be compared with experiment and with QM\@.  Models will generically also have other parameters, located between the designated regions, but these are not necessary for the main argument.  Also
not included in $\lambda$ are non-spacetime-based entities, such as multiparticle wavefunctions, which may be utilized in some models. 

From the assumption of {\bf Local Causality}, specifically from {\bf BSA}, \eq{eq:BL}, it follows that 
\begin{equation}
\label{eq:BL1}
P_{a,b,c}(A|\lambda,B) = P_{a,c}(A| \lambda),
\end{equation} 
because $\bf{\Lambda}$ screens $\bf{1}$ from $\bf{2}$, in the sense that the necessary $S''$ region can be chosen to be fully contained in $\bf{\Lambda}$.  Similarly, $P_{a,b,c}(B|\lambda,A) = P_{b,c}(B|\lambda)$.  It also follows from {\bf NFID} that any model-generated probabilities of $\lambda$ must be independent of the settings $a$ and $b$, because those settings lie in the future of $\lambda$. In equation form, following (\ref{eq:SI2}), this reads
\begin{equation}
\label{eq:SI}
P_{a,b,c}(\lambda)=P_c(\lambda).
\end{equation}
This is often known as ``measurement independence,'' a term that unfortunately obscures the input nature of the measurement settings.  It is clearer to call this condition {\bf $\bm{\lambda}$-independence}, as the equation specifies that $\lambda$ is independent of the inputs $a,b$, via a direct application of {\bf NFID}.%
\footnote{
\label{fn:superdet}%
A different perspective results if one denies free-input-parameter status to the measurement settings, treating $a$ and $b$ as stochastic variables instead of inputs.  This is the ``superdeterministic'' scenario, to be discussed in Section~\ref{sec:superdeterminism}, which allows a version of Eqn.~(\ref{eq:SI}) to be considered as a ``no conspiracy'', ``freedom of choice'', or even a ``free will'' condition.}

Basic probability theory provides a product expression for the joint conditional probability: 
$P_{a,b,c}(A,B|\lambda)=P_{a,b,c}(A|B,\lambda)P_{a,b,c}(B|\lambda)$.  Since $\lambda$ is hidden, the observable joint probability is found by summing or integrating this over all possible values of $\lambda$.  Applying {\bf BSA} and {\bf NFID}, by substituting Eqns.~(\ref{eq:BL1}) and (\ref{eq:SI}), yields Bell's ``Separability Condition'':
\begin{equation}
\label{eq:Sep}
P_{a,b,c}(A,B) = \int d\lambda \; P_{c}(\lambda) P_{a,c}(A|\lambda) P_{b,c}(B|\lambda),
\end{equation}
where the integral is understood as a sum if $\lambda$ is discrete, or a functional integral if $\lambda$ is a function.  This must hold for every applicable model respecting {\bf Local Causality}.

\subsubsection{A Bell inequality}
\label{sec:CHSH}

From Bell's Separability Condition, Eqn.~(\ref{eq:Sep}), one can derive the CHSH inequality \cite{clauser1969}, a generalized version\footnote{See, \textit{e.g.}, \textcite{bell1971} for details.} of Bell's original inequality \cite{bell1964}.  It applies to models for which the output parameters in regions $\bf{1}$ and $\bf{2}$, \textit{i.e.}, the outcomes $A$ and $B$, have two possible values.\footnote{
The proof can be generalized to measurements with continuous results, provided their ranges are restricted, $|A|,|B| \leq 1$.}  
Assigning $\pm 1$ to the outcome values on each side, the product $AB$ must then also be $\pm 1$.  Its expectation value for given inputs, \textit{i.e.}, the correlator of the outcomes, is denoted:
\begin{equation}
\label{eq:correlation}
\langle A B \rangle_{a,b,c}\equiv \sum_{A,B} A \, B \, P_{a,b,c}(A,B).
\end{equation}
The CHSH inequality restricts the values of a combination of correlators, which involves two of the possible settings of the input parameter $a$ in region $\bf{1}$, labelled $a$ and $a'$, and two possibilities for the input setting in region $\bf{2}$, labelled $b$ and $b'$.  The source input setting $c$ is held constant while the four possible combinations of inputs are manipulated, and will be suppressed from here on (we will later consider only particular Bell states, for which only one value of $c$ is relevant).  
It is customary to transfer the primes to the $A$ and $B$ parameters, so that, \emph{e.g.}, $\langle A' B \rangle$ stands for $\langle  A B \rangle_{a',b}$.

With this notation, the CHSH inequality concerns the combination
$\langle A B \rangle + \langle A' B \rangle + \langle A B' \rangle - \langle A' B' \rangle$.  It is easiest to evaluate this combination by sampling the probability distributions in Eqn.~(\ref{eq:Sep}) many ($N$) times, in the style of a Monte-Carlo simulation.%
\footnote{The proof here follows \textcite{peres1978}.  The discussion of the mathematical model rather than the modelled physical experiments avoids the need for any additional assumptions, such as ``counterfactual definiteness'' (the assumption that when $A_n$ is measured, it is legitimate to discuss $A'_n$ as well).}
Denoting the $n$th value sampled from $P(\lambda)$ by $\lambda_n$, we have $A_n$ sampled from $P_a(A|\lambda_n)$ and $A'_n$ from $P_{a'}(A|\lambda_n)$, and similarly for $B_n, B'_n$.
The large $N$ limit is implied, so that, \emph{e.g.}, 
$\langle A' B \rangle = \frac{1}{N} \sum_n A'_n B_n \,$.
The above combination of correlators is then obtained by averaging over 
$(A_n + A'_n) B_n + (A_n - A'_n) B'_n$, and it follows from $A_n, A'_n = \pm 1$ that for each $n$ one of the parentheses must vanish.  The averaged combination therefore is of absolute magnitude 2 for each $n$, and the combination of correlators cannot be larger in magnitude:
\begin{equation}
\label{eq:CHSH}
\left | \langle A B \rangle + \langle A' B \rangle + \langle A B' \rangle - \langle A' B' \rangle \right | \le 2.
\end{equation}
This is the CHSH inequality.

\subsubsection{Contradiction with QM and experiment}
\label{sec:contra}

When the Bell inequalities were first derived, they were shown to be in conflict with the predictions of QM\@.  Now, they are known to be in direct conflict with actual experiments \cite{hensen2015,giustina2015,shalm2015,rosenfeld2017}, independent of the formalism of QM, demonstrating the failure of {\bf Local Causality}.

It is simple to demonstrate that at least some QM predictions violate the CHSH inequality, Eqn.~(\ref{eq:CHSH}).  Consider two photons entangled in a spin-zero Bell state, as in several of the early experiments \cite{clauser1978,aspect1981}.  (Equivalently, two spin-$1/2$ particles can be analyzed.)  Suppose each photon encounters a polarizing beamsplitter, with outputs directed onto two single-photon detectors.  The two beamsplitters are aligned at angles $a$ and $b$ in regions $\bf{1}$ and $\bf{2}$ respectively (these are the measurement settings, defined modulo $\pi$).  For the outcome parameters $A$ and $B$, assign a value of $+1$ when the detectors imply a measured polarization aligned with the setting, and $-1$ for a measurement of the perpendicular polarization.  The predictions of QM are then given by the probabilities
\begin{equation}
\label{eq:QMsinglet}
p_{a,b}(A,B) = \frac{1}{4} \left[ 1 + A B \cos(2a-2b) \right].
\end{equation}
The expectation value of the product $AB$ is therefore $\langle A B \rangle=\cos(2a-2b)$.

For certain combinations of settings, this violates the CHSH inequality by a wide margin. The largest violation obtains for $a=0, a'=\frac{\pi}{4}, b=-b'=\frac{\pi}{8}$, for which the left hand side of (\ref{eq:CHSH}) is $2\sqrt{2}$ (each of the four terms contributes $+1/\sqrt{2}$).%
\footnote{\label{fn:Tsirelson}%
\textcite{cirelson1980} has shown that this is the maximal value achievable in QM, while \textcite{popescu1994} have devised a synthetic model which reaches even higher values, up to $4$, the maximum possible.}
These non-classical correlations between the two photons served historically as an early and striking example of the much wider family of phenomena associated with quantum entanglement [see, \emph{e.g.}, \textcite{brunner2014,streltsov2017}].

The observed violations of the inequalities are by impressive margins, greatly exceeding the experimental accuracy.  Indeed, as an empirical test of a mathematical model or a class of models, the confidence with which the CHSH inequality is rejected approaches the certainty of a mathematical proof.  For example, the experimental results of \textcite{giustina2015} boast a value less than $3.7 \cdot 10^{-31}$ for the probability that the results could be obtained under the assumption of {\bf Local Causality}, according to the standard statistical analysis.  Furthermore, this result belongs to the recent generation of ``loophole-free'' experiments (those cited above), which are free from all of the simplifying assumptions which were necessary for Bell tests with earlier technology.  The observations not only violate the CHSH inequality---the quantitative results follow the predictions of QM in fine detail.  We now turn to models that can be consistent with these experiments, Bell's Theorem notwithstanding.

\section{Implications}
\label{sec:aftermath}

The upshot of Bell's Theorem is that there is no longer any hope of finding a reformulation of QM which respects {\bf Local Causality}.  But the use of spacetime-based parameters has not been ruled out altogether, and the motivations for using them remain intact.  Furthermore, given such parameters, there are still live options for saving {\bf CA} ({\bf Continuous Action}), the generalized form of ``locality'' defined in Section II.B.1.  In this sense, Bell's Theorem does {\em not} necessarily imply unmediated action-at-a-distance.

The rest of this Colloquium is dedicated to an analysis of the possibility of reformulating QM in a ``locally-mediated'' manner, consistent with both {\bf CA} and Lorentz covariance.  Recall that a ``reformulation'' here means a model in {\bf agreement} with QM, with the same inputs $I$, the same outputs $O$, and the same model-generated joint probabilities $P_I(O)$.  In preparation for this, the first subsection below proposes a categorization scheme for all models in {\bf agreement} with QM, and the second clarifies relevant issues of causation and signaling.

\subsection{Categories of Bell-compatible reformulations of QM}
\label{sec:categories}

As stated, Bell's Theorem dictates that no model in {\bf agreement} with QM can respect {\bf Local Causality}, which is the conjunction of two assumptions: {\bf NFID} ({\bf No Future-Input Dependence}) and {\bf BSA} ({\bf Bell's Screening Assumption}).  Reformulations of QM must thus violate at least one of these in some non-trivial manner, such that the CHSH inequality can also be violated.  Of these two assumptions, we argue that the primary one for categorization purposes should be {\bf NFID}, because it is often taken for granted, and because the motivation for {\bf BSA} in Section~II included {\bf NFID}, as depicted in Figure 1.  A useful secondary categorization is the {\bf CA} condition, indicating whether or not action-at-a-distance is implied by a given model.  

Bell's Theorem thus requires all models in {\bf agreement} with QM to fall in one of the following categories:
\begin{itemize}
\item Type {\bf I}: Respect {\bf NFID}
    \begin{itemize}
    \item Type {\bf IA}: Respect {\bf CA} (Must Violate {\bf BSA})
    \item Type {\bf IB}: Violate {\bf CA}
    \end{itemize}
\item Type {\bf II}: Violate {\bf NFID}
    \begin{itemize}
    \item Type {\bf IIA}: Respect {\bf CA} (May Violate {\bf BSA})
    \item Type {\bf IIB}: Violate {\bf CA}
    \end{itemize}
\end{itemize}
(Models which violate {\bf CA} necessarily also violate {\bf BSA}.)  For convenience, the different Types of models are also identified in Table I.

\begin{table}[b]
\label{Table:Types}
\caption{Categories of possible reformulations of QM (and the sections in which they will be discussed).  The columns identify whether or not a model conforms with the CA locality condition, and the rows refer to the NFID arrow-of-time condition.  Bell's Theorem rules out the subset of Type~{\bf IA} models which conform also to the stricter {\bf BSA} (Bell's Screening Assumption) locality rule (see Fig.~1 above).  In the following, we will focus on locally-mediated models, which are of Type~{\bf IIA}.}
\centerline{\includegraphics[width=.7\textwidth]{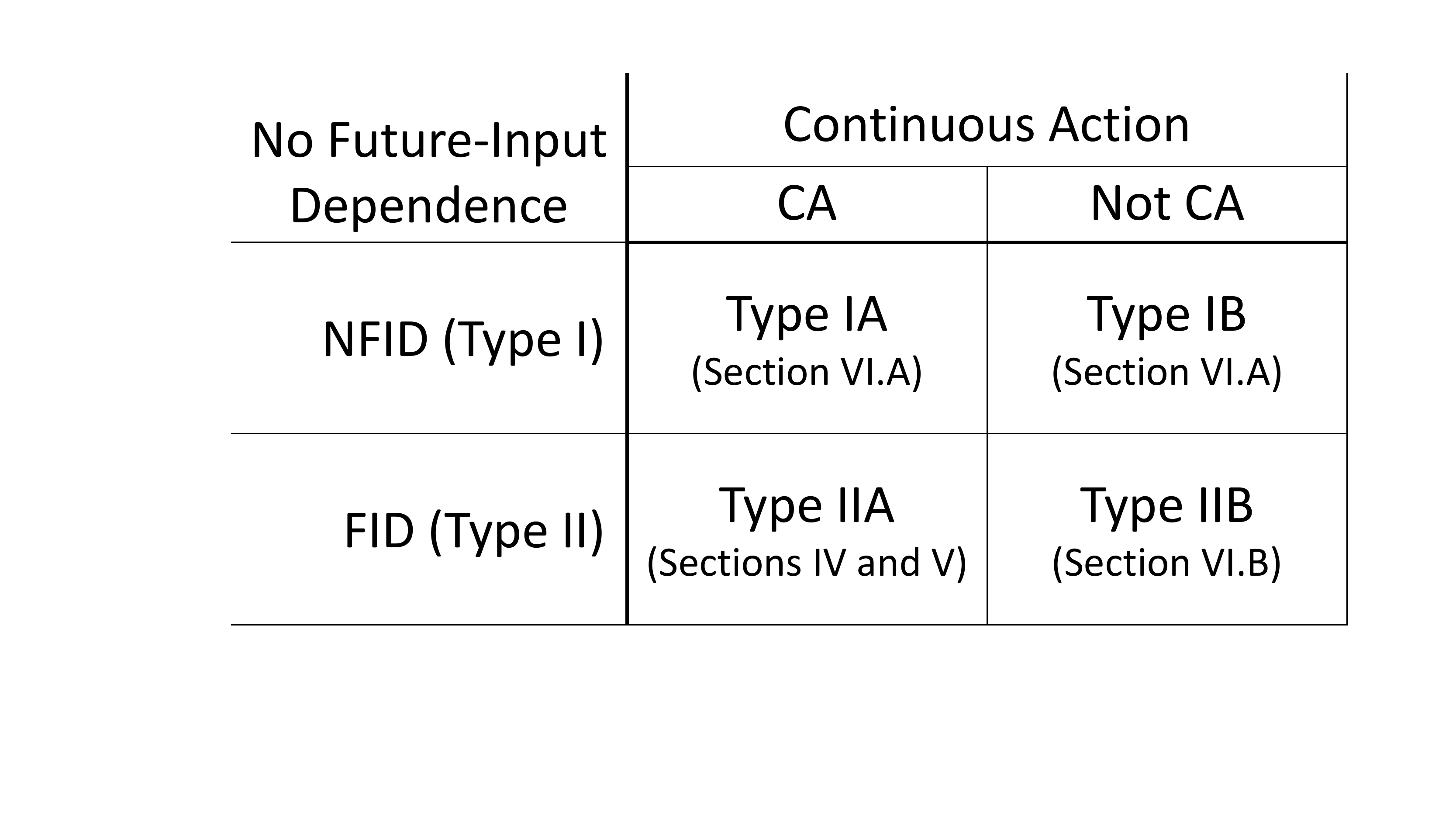}}
\end{table}

From the definition of {\bf NFID}, Type {\bf I} models allow for the calculation of all spacetime-based parameters in temporal order, using inputs that enter into the calculation in that same order.  But because of the necessary {\bf BSA} violation, such models cannot adhere to the light-cone-constrained Cauchy problem typically found in classical physics.  

Type {\bf IA} models would have to avoid {\bf CA}-violation using faster-than-light mediators, bypassing the screening region $\bf{S''}$ of Figure 1(c) but passing through the larger screening region $\bf{S'}$ of Figure 1(b).  Such models have not been formally developed, but have been promoted by various authors, including \textcite{bell1981} himself.  In order for this not to violate {\bf NFID} in a different reference frame, one might propose a special frame in which the model uniquely applies, at the expense of Lorentz covariance.  
Moreover, to maintain {\bf agreement} with QM in all cases, it is necessary that the mediating signal should always pass through $\bf{S'}$ even if this region is blocked, say by a brick wall.  For these reasons, we judge such models to be of less interest, and our use of the term ``locally-mediated'' will exclude such faster-than-light unblockable mediators.

The Type {\bf IB} category includes the standard Schr\"odinger-picture QM itself, as well as less conventional approaches such as de Broglie-Bohm guiding-waves \cite{bohm1952}.  Such models utilize \emph{mathematical} intermediaries $R$ to connect distant spacetime-based parameters.  Recall that the model parameters ($I$, $Q$) are defined as associated with particular places and times.  Values in $R$ might be associated with multiple spacetime locations in some non-separable manner, and do not generally have a form such as $R(\bm{x},t)$.%
\footnote{The efforts of \textcite{norsen2010,stoica2019} aim to  overcome this, which could lead to Type {\bf IA} models.}
The most prominent example of a parameter $R$ is the many-body wavefunction, which for an entanglement setup is defined on configuration space.  Such an account involves no parameters ${\lambda}$ in the relevant space-time regions, directly violating both {\bf BSA} and {\bf CA} [Eqn.~(\ref{eq:Sep}) becomes a simple product, with ${\lambda}$ representing a constant, the empty set].  The role of the wavefunction in producing the predictions of QM might be described as an abstract mathematical object connecting events in spacetime.

Type {\bf II} models violate the {\bf NFID} assumption, so they are not temporally-sequential calculations.  It is natural to call such models {\bf Future-Input Dependent}, or {\bf FID} models.  With well-known examples such as the stationary action principle, it is clear that {\bf FID} models cannot be trivially dismissed, and yet they are rarely brought up in discussions of Bell's Theorem.%
\footnote{One factor which surely contributed to this is that Bell himself did not mention the FID possibility in any of his publications \cite{bell2004}, see Section~\ref{sec:cause_effect}.  This omission continued in several major reviews, \textcite{goldstein2011} and \textcite{shimony2017}, although the latter has recently been updated with a recognition of retrocausation \cite{myrvold2019}.}

Type {\bf IIA} models are particularly interesting because they do not involve action-at-a-distance, in the sense that the screening condition of {\bf CA} is respected.  Furthermore, since {\bf NFID} is already violated, Bell's Theorem does not rule out the possibility of even retaining the stricter {\bf BSA} locality condition.  As we shall see, in a Type {\bf IIA} model, the mediation between spacelike regions can take place entirely on timelike worldlines.  Instead of the unblockable faster-than-light mediators of Type {\bf IA} models, all relevant parameters in Type {\bf IIA} models can be associated with the actual particle histories, allowing Lorentz covariance to be preserved.  This in turn means that it is possible to build models without any abstract mathematical structures $R$ ``mediating'' events in conventional spacetime.  Of course, one could still use such structures if desired -- say, by retaining the conventional QM configuration-space wavefunction in a model.  This would fall into the category of Type {\bf IIB} models which violate the intuitive {\bf NFID} condition while restoring neither {\bf CA} nor {\bf BSA}.

The next two sections will be devoted, respectively, to a review of the specific achievements of the Type {\bf IIA} toy-models which have already been developed, and to a discussion of the drawbacks and promise of this category of models.  The other categories, as well as approaches which do not fall within the framework used here, will be discussed in Section~\ref{sec:alternatives}.  Before this review, some additional clarifications are necessary, to which we turn next.

\subsection{Causality and locality}
\label{sec:more_assumptions}

The failure of {\bf Local Causality} implied by Bell's Theorem leads naturally to the question: In what sense, if at all, does {\bf Local Causality} correspond to assumptions of locality and causality?  Before continuing, it is necessary to clarify these issues.

\subsubsection{Cause and effect}
\label{sec:cause_effect}

The definition of {\bf NFID} in Section~\ref{sec:assumptions} uses the distinction between input- and non-input-parameters, rather than the words ``cause'' and ``effect.''  Nevertheless, the {\bf NFID} condition is closely related to a definition of causality which arises naturally within the modern account of ``interventionist'' causation, where causes are identified as interventions \cite{pearl2009,woodward2005}.    If the input parameters in question are deemed to be controllable parameters, then it is appropriate to identify them as causes, according to this account.

QM itself clearly adopts this connection between inputs and controllable parameters: the mathematical formalism of QM is a procedure for making operational predictions for observations, given the values of the controllable inputs.  As our goal is to discuss models in {\bf agreement} with QM, it is natural for us to adopt this approach.  Such models limit the inputs $I$ to the parameters that QM tells us can be externally controlled. 

Given this connection between ``controllable inputs'' and ``causes'', one can identify different possible causal structures.  In models that respect {\bf NFID}, non-input parameters are typically functionally dependent on past inputs, but are always functionally \textit{independent} of future inputs. This ``forward-causal'' structure is clearly what Bell had in mind when he used the terms ``causality'' and ``causal structure,'' with the controllable inputs called ``free variables'' or ``free elements'' \cite{bell1977,bell1990}.\footnote{
There is an early exception: \textcite{bell1964} used ``causality'' to imply ``complete causality,'' \emph{i.e.}, determinism.}

{\bf FID} models, on the other hand, do not have a forward-causal structure.  In other words, they cannot generally compute a given parameter $q(t')$ (or its probability distribution) without specifying certain inputs in the future of $t'$.  In the framework of interventionist causality, if those future inputs are controllable, the {\bf FID} models are ``retrocausal''.%
\footnote{The word ``retrocausal'' conventionally implies there are \textit{some} future causes of \textit{some} past parameters, not a purely-reverse-causal structure.}

Some {\bf FID} models, such as classical action principles, are not retrocausal.  In those cases, the final boundary constraints are required mathematical inputs, but not {\em controllable} inputs, and so are not considered causes.  Analysis of the causal structure of such a theory requires inverting the functional relation between some of the inputs and some of the outputs, so that a different model is obtained---a model in which all inputs are controllable.  Although it makes sense to refer to the action principle itself as a reformulation of Newton's equations, it is only after this inversion that one obtains a model fully in {\bf agreement} with the standard operational description of classical mechanics, which uses the controllable initial conditions as inputs.

At the time of Bell's work, the interventionist approach to causation had not yet been well-developed.  An older approach was taken for granted, dictating that if two parameters exhibit cause-effect correlations, it is appropriate to refer to the one earlier in time as a cause, and the later one as an effect, regardless of which one can be externally controlled.

This is one topic where one's definition of causation directly impacts the types of mathematical models that one views as acceptable.  Applied to the {\bf $\bm{\lambda}$-independence} condition, any violation of \eq{eq:SI} would be viewed as retrocausal in the framework of interventionist causation, an instance of {\bf FID}.  But if one instead assumed that $\lambda$ was the cause of the settings $a,b$, because $\lambda$ occurs before $a,b$ were chosen, one would have to conclude that the settings were \emph{effects}, and could not be treated as free inputs (see footnote~\ref{fn:superdet} above and Section~\ref{sec:superdeterminism} below).  The model would then not be in {\bf agreement} with QM.

\subsubsection{Signals}

Just as QM restricts the inputs $I$ to be {\em controllable}, it also specifies that the outputs $O$ are {\em observable}.  If $I$ is controllable and $O$ is observable, $P_I(O)$ summarizes all possible {\em signals}.  And as QM does not allow signals to be sent back in time, it follows that for models in {\bf agreement} with QM the outputs $O$ cannot depend on future inputs.  We shall call this requirement {\bf signal causality}, or explicitly,
\begin{equation}
\label{eq:SC}
P_I(O')=P'_{I'}(O'),
\end{equation} 
where the primed sets of parameters are all those associated with times up to $t'$, as in the similar Eqn.~(\ref{eq:NFID}).  

Comparison with Eqn.~(\ref{eq:NFID}) indicates that any violation of {\bf NFID} in a model in {\bf agreement} with QM must be at the level of unobservable (hidden) parameters $U$ in $Q$.  Such an FID model would be retrocausal (at a hidden level), but would not violate {\bf signal causality}.%
\footnote{
If one demanded not only that ``causes'' are identified with controllable inputs but also that ``effects'' are identified with observable outputs, one would be led to take \eq{eq:SC} as representing the causal arrow of time.  However, the term retrocausal in the literature does not signify violations of {\bf signal causality}.  We use the more technical term {\bf NFID}, which explicitly focuses on inputs, in order to minimize confusion.}

Motivated by special relativity, it is natural to formulate a stronger restriction on signaling.  This condition, called {\bf signal locality}, limits signals to traveling no faster than light, so that signals associated with a particular controlled input are limited to outputs in its future lightcone.  For outputs $O_1$ localized in region $\bf{1}$, the relevant inputs $I''$ should thus lie in the past lightcone of $\bf{1}$, and the {\bf signal locality} requirement corresponds to the existence of a restricted model $P''$ such that
\begin{equation}
\label{eq:SL}
P_I(O_1)=P''_{I''}(O_1).
\end{equation}
This condition also holds in QM, and must thus be maintained for any model in {\bf agreement} with QM.

As indicated in the introduction, these signal-based definitions of ``locality'' and ``causality'' are operational, in the sense of involving only controllable inputs and observable outputs.  Bell's Theorem states that models in {\bf agreement} with QM must violate either a distinct notion of ``locality'' ({\bf BSA}), or a distinct notion of ``causality'' ({\bf NFID}) which are not defined operationally, as they refer to hidden model parameters, not signals.   Because of these different definitions, models can be local (or causal) in one sense, but not in another.%
\footnote{The literature on Bell's Theorem involves quite a few additional ``locality'' conditions [see, \emph{e.g.}, \textcite{wiseman2014}], but these are not needed for the present discussion.}

\section{Locally-Mediated Models of Entanglement (Type {\bf IIA}) }
\label{sec:TypeIIA}

This section will discuss reformulations of QM which fall in Type {\bf IIA}, meaning that they are ``locally mediated'' as discussed above.  These models conform to {\bf CA} ({\bf Continuous Action}) and are {\bf FID} ({\bf Future-Input Dependent}), allowing for compatibility with Bell's Theorem without a necessary conflict with Lorentz covariance.  As noted, such models are underrepresented in the literature on Bell's Theorem, so this section and the next will provide a rather thorough discussion.  

The essential strategy behind Type {\bf IIA} models of entanglement is to allow a violation of {\bf $\bm{\lambda}$-independence}, Eqn.~(\ref{eq:SI}), such that $P_{a,b}(\lambda)$ is not independent of the input settings $a,b$.  The relevant $\lambda$ lies in the past light-cones of $a,b$, so that such models are technically ``retrocausal'' as defined in Section~\ref{sec:cause_effect}.  But as noted there, if {\bf agreement} with QM is to be maintained, any correlations with future settings must be sequestered in hidden variables, not observable outputs.  By restricting attention to models in {\bf agreement} with QM, there is no possibility of signals being sent back in time, and thus no concern of generating paradoxes.  These and other concerns with such models will be further discussed in the next section.  

The promise of Type {\bf IIA} models is that, in any given case, there exist parameters $\lambda$ that can act as local mediators of the actual correlations.  It is always simple to find a distribution of shared parameters $\lambda$ that will produce a given correlation for particular measurement settings; Bell showed that the problem was getting the same $P(\lambda)$ distribution to consistently work for \emph{all} measurement settings.  But for models $P_{a,b}(\lambda)$ where the distributions can be different for different settings, Bell's consistency problem disappears. This means that it is possible to retain {\bf BSA} ({\bf Bell's Screening Assumption}) in some {\bf FID} models, or at least the weaker locality condition {\bf CA}.  

At the current stage of development of Type {\bf IIA} models, there are none which are applicable to a wide range of quantum phenomena.  Existing models aim at reproducing merely the known correlations for the Bell state, Eqn.~(\ref{eq:QMsinglet}).  Several will be presented below, with schematic models in the fist subsection, and a model providing a more detailed description in the second.

\subsection{Schematic models}
\label{sec:PoP}

Although the idea of using future-input dependence to explain entanglement had been around for a long time \cite{costa1953,costa1977,costa1979, cramer1980,pegg1982,sutherland1983,price1984,price1997}, explicit Type {\bf IIA} mathematical models of entanglement have appeared in the literature mostly in the last decade.  One notable exception is \textcite{pegg1982}, a description which could be simplified%
\footnote{One complication is that in its original form, the intermediate state appears to be {\em output}-dependent, rather than dependent on the future input setting.}
and expressed in a manner quite similar to that of \textcite{argaman2010}, the model presented next.

\subsubsection{A simplistic model}

Consider again the correlations between the polarizations of a pair of entangled photons.  Using the terminology of Section~\ref{sec:contra}, where $a$ and $b$ represent the angle settings of polarizers, the spin-zero Bell state correlations can be obtained from the following toy-model.  First, take the two photons to both be initially polarized at an angle $\lambda$, distributed according to 
\begin{equation}
\label{eq:Argaman}
\begin{split}
P_{a,b}
(\lambda) = \frac{1}{4} \left\{ \frac{}{} \delta (a-\lambda) +
\delta \left(a+\frac{\pi}{2}-\lambda \right) + 
\right. \quad \\ \left. \;\;\; 
\delta(b-\lambda) + \delta \left(b+\frac{\pi}{2}-\lambda \right)  \right\}.
\end{split}
\end{equation}
Here $\lambda \in [0,\pi)$ and the $\delta$-functions are modulo $\pi$.  In this model, the initial polarization $\lambda$ is thus somehow constrained by the future settings to be either $a$, $a+\pi/2$, $b$ or $b+\pi/2$ with equal probabilities, \emph{i.e.}, to be aligned with one of the detectors.  

Next, apply Malus' law to obtain the results of the single-photon measurements, $A$ and $B$ [\emph{i.e.}, $P_a(A=1|\lambda)=\cos^2(a-\lambda)$, etc.]. Combining these, using Eqn.~(\ref{eq:Sep}), reproduces the QM probabilities for the spin-zero Bell state, \eq{eq:QMsinglet}.  While this model is clearly very schematic, it demonstrates that only mediation along the spacetime paths of the particles is required.  

\subsubsection{The Hall model}
\label{sec:Hall}

A number of additional Type {\bf IIA} schematic models follow a similar strategy.  They consist of two components: (i) a specification of the sample space of the hidden variables and their distributions $P_{a,b}(\lambda)$, and (ii) models for the measurement outcomes, $p_{a,[\lambda]}(A)$ and $p_{b,[\lambda]}(B)$, such that the combination of (i) and (ii) per Eqn.~(\ref{eq:Sep}) is in {\bf agreement} with QM for a specific setup of interest.  (The notation $[\lambda]$ emphasizes that while $\lambda$ is an input to the second component, it is not an external input.)

For want of space, we will provide the details of just one additional example, that of \textcite{hall2010}.  The version adapted to photon polarizations \cite{argaman2018} has:
\begin{equation}
\label{eq:Hall}
P_{a,b}
(\lambda) = \frac{1}{\pi} \, 
\frac{1 + \grave A \grave B \cos(2a-2b)}{1 + \grave A \grave B (1-z)} ,
\end{equation}
where $\grave A = {\rm sign}[\cos(2a-2\lambda)]$, $\grave B = {\rm sign}[\cos(2b-2\lambda)]$ and $z=\frac{2}{\pi} |2a-2b|$ are abbreviations.  This model is deterministic in the sense that $A$ is fully determined by $a$ and $\lambda$ through $p_{a,[\lambda]}(A)=\delta_{A,\grave A}$, with the same expression relating $B$ to $b$ and $\lambda$.  It reproduces the results of QM for the Bell state, Eqn.~(\ref{eq:QMsinglet}).

Here, knowledge of $\lambda$ provides only a very rough idea of what $a$ and $b$ are.  When properly quantified, the information about $a$ and $b$ which can be gleaned from the past parameter $\lambda$ amounts to less than 0.07 bits per entangled pair \cite{hall2016}.  In this sense, one may view the toy-model of Eqn.~(\ref{eq:Hall}) as a dramatic improvement over that of Eqn.~(\ref{eq:Argaman}).

\subsubsection{Additional toy-models}
\label{sec:more_toys}

A large number of additional schematic entanglement models exist in the literature, the majority of which are Type~{\bf IB} models.  The original \textcite{bell1964} work contained such a model for ``illustration'' purposes, and many others were developed over the years, relying on different ``resources'': communication, shared randomness and/or nonlocal boxes [see, \emph{e.g.}, \textcite{degorre2005}, and references therein].  Each of these models proposes a novel distribution which may be denoted by $P_{a,b}(\lambda)$, and a way in which that $\lambda$ generates the output statistics of QM, per steps (i) and (ii) above.  Simply associating $\lambda$ with the world-lines of the entangled particles, rather than with the time of the measurements, can then change the Type~{\bf IB} model into a Type~{\bf IIA} model.  An example is given by \textcite{barrett2011}, who modified the model of \textcite{toner2003,degorre2005} by ``moving'' $\lambda$ to the past.

Other changes in the spacetime location of $\lambda$ can affect the assessment of the {\bf NFID} and {\bf CA} conditions, leading to a new model of a different Type, even if the distribution $P_{a,b}(\lambda)$ is unchanged.  For Type~{\bf IIA} models, taking $\lambda$ to be associated with the emission event at the source but not with the particle worldlines---as arguably done in the machine-learning-generated models of \textcite{weinstein2017,weinstein2018}---formally results in a {\bf Type IIB} model, with no local mediators.  But such models are easily transformed back into {\bf Type IIA}, simply by copying $\lambda$ onto mediators along both worldlines.  Alternatively, $\lambda$ might be associated with the time of the measurements, rather than the emission or the worldlines; resulting in a model of {\bf Type IB}.  Reinterpreting Eqn.~(\ref{eq:Argaman}) in this manner leads to precisely the model of \textcite{dilorenzo2012}.  Yet another example is provided by the work of \textcite{sen2018}, who began with the model of \textcite{brans1988} (itself obtained by associating the parameters of standard QM with the past) and explicitly transformed it into a Bohmian-style {\bf FID} model.

The schematic Type {\bf IIA} models above show both promise and limitations.  On the positive side, they all serve as proof-of-principle examples, indicating that Bell inequalities can be violated without also introducing action-at-a-distance, and they provide a variety of points of departure for future development.  With the mediating parameters $\lambda$ associated with the particle worldlines, other advantages quickly become evident.  For example, a recent application \cite{sen2019} of a {\bf FID} model to entanglement in accelerating reference frames indicates a nearly-trivial reconciliation of quantum phenomena and general relativity, for a case that is even quite difficult for quantum field theory.  

On the negative side, however, these models all simply assert the connection between the settings and $\lambda$, without a proposed mechanism or explanation.  One natural justification for such a connection would be an appeal to time symmetry: one could argue that the symmetry exhibited by micro-scale phenomena implies an equal role for both past and future.  This would make the future settings $(a,b)$ just as important as the initial state preparation $(c)$ when modeling $\lambda$.  But this justification seems inapplicable because these schematic models do not possess time symmetry in any sense.  We now turn to a Bell-compatible {\bf FID} reformulation which restores microscopic time symmetry, and does so in a manner that provides an account of both the $P_{a,b}(\lambda)$ distribution and the outcome probabilities.

\subsection{The Schulman L\'evy-flight model}
\label{sec:Schulman}

Conventional QM is typically viewed as time-symmetric, but its intermediate calculations are notably time-asymmetric.  For example, consider the polarization of a photon which is known to have passed through two consecutive polarizers set at angles $\theta_1$ and $\theta_2$.  The conventional description associates the angle $\theta_1$ with the polarization of the photon between the two polarizers, but time-reversal symmetry implies that $\theta_2$ should be just as relevant to the intermediate description.  Any time-symmetric account of the intermediate photon should therefore take both angles into account, and would be a Type {\bf IIA} model.

Such a time-symmetric model has been developed by \textcite{schulman1997,schulman2012}, using a time-varying polarization angle $q(t)$.\footnote{
Schulman's discussion of spin-$1/2$ particles is here adapted to photons.}  
Schulman considered the possibility that $q(t)$ could be perturbed by microscopic rotations $dq$ (``kicks'') so that $q(t)$ evolves from $\theta_1$ to $\theta_2$ (or $\theta_2 + n \pi$) between the polarizers, without requiring a collapse at the last instant.  If the magnitude of each microscopic kick is normally distributed (or has a finite second moment) one would obtain diffusive behavior, which is inappropriate.  However, if $q(t)$ describes a L\'evy flight, \emph{e.g.}, if the magnitudes of the kicks are distributed according to the Cauchy (Lorentzian) distribution, 
$\propto d\gamma / [{(dq)^2+(d\gamma)^2}]$ with a small width $d\gamma$, the net rotation $\Delta q$ has a similar probability distribution:
\begin{equation}
\label{eq:SProb}
P(\Delta q) = \frac{1}{\pi} \frac{\gamma}{(\Delta q)^2+\gamma^2},
\end{equation}
where $\gamma$ is the sum of all the $d\gamma$ widths of all the kicks along the path.

With $q(t)$ constrained to $\theta_1$ at the time of the initial polarizer, $t_{\rm i}$, and to $\theta_2$ at $t_{\rm f}$, the final time, $q(t)$ provides an appealing time-symmetric description of the dynamics (constrained by initial and final boundaries).  Moreover, and this is the main point of Schulman's derivation, the model correctly predicts the outcome probabilities for a single photon in the limit $\gamma\to 0$, if the measurement acts as a boundary constraint corresponding to discrete possibilities, requiring the photon polarization to either be aligned or perpendicular to the polarizer angle (either $\theta_2$ or $\theta_2 + \pi/2$).  Adding all the equivalent contributions corresponding to $\theta_2 + n\pi$ per Eqn.~(\ref{eq:SProb}) gives a result $\propto 1/\sin^2(\theta_1-\theta_2)$. Comparing this to the other possible outcome, summing over $\theta_2 + (n+\frac{1}{2})\pi$, reproduces Malus' law upon normalization: the probability for a photon of initial polarization $\theta_1$ to align with a polarizer oriented at $\theta_2$ is $\cos^2(\theta_1-\theta_2)$.  A detailed derivation can be found in the Appendix.

Note that for small $\gamma$ the path $q(t)$ is very close to being a constant, but the initial and final requirements enforce at least one significant ``kick'', with a distribution $\propto d\gamma / (dq)^2$.  In the $\gamma\to 0$ limit, paths with a single kick dominate.  There is thus an event which corresponds to ``collapse'' in this description (unless $\theta_1=\theta_2$ or $\theta_1=\theta_2+\pi/2$), but it happens at an arbitrary time between preparation and measurement, rather than at the time of the measurement, and thus respects time symmetry.

This model can be trivially extended to the case of two maximally entangled photons, by combining two copies of the single-particle model, $q_1(t)$ and $q_2(t)$, and constraining their unknown initial polarization angles to be identical, $q_1(t_{\rm i}) = q_2(t_{\rm i})$, \cite{wharton2014, almada2016}.  Identifying this initial polarization as the hidden parameter $\lambda$ reproduces precisely the probability distribution of the simplistic toy-model above, \eq{eq:Argaman}.  This follows because the overwhelmingly most probable scenario is to have only one significant kick in the combination of the two paths, and this in turn requires $\lambda$ to match one of the two future settings.  

In this model, the screening region $S''$ from Figure 1(c) contains the parameters $q_1(t)$.  No inputs on the other arm of the experiment can affect the probability of the measured outcome $q_1(t_{\rm f})$ without also affecting the earlier values $q_1(t)$, conforming to {\bf BSA}.  (The earlier schematic models also respect {\bf BSA} for similar reasons.)
The mechanism by which the correlations are enforced is {{\bf NFID}-violating}: the future settings $(a,b)$ constrain the full histories $q_1(t)$ and $q_2(t)$, including the possible initial value of the hidden parameter, $\lambda$.  This explicitly violates {\bf NFID} and Eqn.~(\ref{eq:SI}), violating {\bf Local Causality}.  All locality conditions from Figure 1 are thus preserved. 

The Schulman Type {\bf IIA} model supplies a future boundary mechanism to explain the future-input dependence [an account missing from component (i) of the above schematic models, as noted there], and the very same mechanism provides the correct outcome probabilities [to explain component (ii)].  Indeed, this two-particle toy-model is currently the most sophisticated example of how a model can yield the correct Bell-state correlations while retaining the {\bf BSA} (or the {\bf CA}) condition of locality. It demonstrates a spacetime-based mediation of the correlations involved in entanglement, via a mechanism that uses the entire history rather than instantaneous ``states''.  By assigning probabilities to histories rather than states, this approach avoids the tension between entanglement and relativistic covariance.  It demonstrates how Type {\bf IIA} models need not conflict with relativity (as noted already in Section~\ref{sec:categories}), because all of the mediation is by parameters that reside on timelike or lightlike worldlines. If the relevant parameters $\lambda$ reside on the classical worldlines of the entangled particles, this looks essentially similar in every reference frame, no matter which particle is measured ``first.''

It is striking that the same set of rules is applicable to both one-photon and two-photon setups, as explained above, and is also valid if additional measurements are considered.  For a single photon, it provides the appropriate Malus-law probabilities for any number of sequential polarization measurements.%
\footnote{It thus provides a ``natural'' mechanism or explanation for violations of Leggett-Garg inequalities \cite{leggett1985}; these ``beyond-Bell'' inequalities facilitate experimental demonstrations of additional surprising quantum phenomena.  Again, the Type {\bf IIA} approach describes a relationship between microscopic hidden variables and macroscopic observable results which appear quite perplexing from a ``macro-realistic'' {\bf NFID}-assuming point of view.}
For the two-photon entanglement setup, as the hidden parameter $\lambda$ is associated with the photon polarization, it is natural to ask whether an additional measurement of this polarization along the path of the photons could shed light on the mechanisms involved.  QM itself describes how this would fail---after such a measurement, the two photons will no longer be entangled.  The Schulman model successfully describes this: the additional measurement would be associated with another boundary constraint, changing the entire history of the experiment, and requiring two ``significant kicks'' instead of one, reproducing again the often-perplexing results of standard QM\@.

The two-particle Schulman model can also be trivially generalized from the spin-zero state to any maximally entangled two-qubit state by performing polarization rotations on one of the two photons.%
\footnote{The strategy of reducing a two-particle entanglement problem to two single-particle problems can be extended to all maximally-entangled bipartite states \cite{wharton2011}.}
Further generalizations to scenarios with several particles \cite{bennett1993,pan1998} might no longer respect {\bf BSA} if some of the correlating parameters are localized on connected zigzagging worldlines (an entanglement-swapping setup), but would continue to respect {\bf CA} and would still be Type {\bf IIA}.  The challenge of extending this type of model to partially-entangled states remains an open problem.

\section{Discussion}
\label{sec:proscons}

In this section, we first address existing criticism of the Type {\bf IIA} approach, and then discuss its potential and directions for future exploration.

\subsection{Objections to Type {\bf II} models}
\label{sec:con_arguments}

Despite the availability of the simple models presented above, much of the contemporary discussion of Bell's Theorem fails to recognize such a possibility.  For example, in the recent round of loophole-free experiments \cite{hensen2015,giustina2015,shalm2015,rosenfeld2017}, not one article mentioned the possibility of Type {\bf II} or {\bf FID} ({\bf Future-Input Dependent}) models.  In the rare case where experimental papers mention a retrocausal option, it is typically relegated to a mere footnote \cite{handsteiner2017,rauch2018}.

With this lack of attention, there are few published concerns about Type {\bf II} models in the recent literature, although a number of ``intuitive'' objections are likely to occur to most physicists upon first encountering these models.  The most common such concerns will be addressed first, followed by a discussion of specific formal arguments which have appeared in the literature.  

\subsubsection{Intuitive objections}

One common objection to {\bf FID} models is that they violate some unwritten principle of ``causality''.  Formalizing this objection is difficult, but one evident concern is that such models might lead to logical difficulties with time-travel paradoxes.  But time-travel paradoxes require communication with the past, with at least some level of observable signal, and this is forbidden in models in {\bf agreement} with QM which conform to {\bf signal causality}, Eqn.~(\ref{eq:SC}) above.  For any {\bf FID} model in {\bf agreement} with QM, the future-input dependence is always at the level of the {\em hidden} parameters, $\lambda$, and as there is no protocol for observing the values of these parameters (without changing the whole setup), such models do not allow retro-signaling.

Another common concern is that {\bf FID} models imply future inputs must ``exist'' to constrain hidden parameters in the past, and some find this block-universe view problematic \cite{sorkin2007,kastner2017}.  But it appears ill-advised to avoid developing a theory for such reasons---it would have been a pity, for example, if Newton were to avoid developing the Law of Universal Gravitation because he perceived its nonlocality to be unacceptable.  Furthermore, treating future events as valid model parameters and analyzing entire spacetime regions ``all at once'' is common in physics, \emph{e.g.}, in general relativity and with Wick rotations.  And in any case, one can always wait until the whole relevant spacetime region is in the past, and perform the model analysis retrospectively.  We set aside this objection as an essentially anthropocentric restriction on mathematical models \cite{wharton2015a}.

As a related objection, some might take the view that because QM conforms to {\bf signal causality}, and so do all other established physical theories, there should never be any reason to consider {\bf FID} reformulations of QM.  However, as we have seen above, the failure of {\bf Local Causality} provides just such a reason.  Bell's Theorem does not formally tell us whether it is the ``locality'' ({\bf BSA}) or ``causality'' ({\bf NFID}) aspects of our models which require adjustment, so we should seriously consider both options rather than simply choosing the one we take to be more plausible.

And again, such a restriction is routinely ignored by physicists in practice.  Histories approaches such as \textcite{griffiths2001}, and path integrals in general, encourage one to consider the past and future together as a single structure, violating the spirit of {\bf NFID}.  In Heisenberg-picture QM, measurement operators are often evolved back in time to the previous measurement. And some analyses of ``delayed-choice'' experiments, such as that of \textcite{bohr1935} briefly described in Section~\ref{sec:interlude}, allow one to make incompatible inferences about past events for different future measurement choices.  If those past events are parameterized, this also violates {\bf NFID}.  

\subsubsection{Formal objections}

An early technical argument against {\bf FID} models is due to \textcite{maudlin1994}.  Adapting it to the above Bell-state setup, consider the case where one measurement is performed early enough so that the result $A$ can be sent ahead of the other particle (say, via a laser signal) to the other measurement device.  This output parameter $A$ could then be used to determine the other setting $b$, via some algorithm $b=f(A)$.  The challenge is one of self-consistency: if one uses a model that requires $b$ as an input to generate $\lambda$, and then uses $\lambda$ to generate the outcomes $A,B$, the function $f(A)$ might be found to disagree with the value of $b$ utilized in the calculation.  This is of particular concern for the schematic models designed with one experiment in mind (such as those in Section~\ref{sec:PoP} above), because this is an essentially different experimental configuration.

But it is unreasonable to expect precisely the \emph{same} model, with the same inputs and outputs, to apply to this new configuration.  In this version of Maudlin's challenge, the setting parameter $b$ is no longer an input to the model (it cannot be freely set), so an analysis of this new experiment would require a Type~{\bf IIA} model of the form $P_a(Q)$, rather than the original $P_{a,b}(Q)$.  The Schulman model of Section~\ref{sec:Schulman} is general enough to handle this new configuration, because the boundary constraints imposed by the future measurements are still enforced in the global solution, no matter whether the settings are free inputs or calculated parameters.  So long as the solution is calculated ``all at once''---assigning probabilities to entire histories rather than states---every intermediate solution is self-consistent, by definition \cite{berkovitz2008,lewis2013,wharton2014}.%
\footnote{In general, the {\bf agreement}-with-QM status of the original $P_{a,b}(Q)$ guarantees through {\bf signal causality} that its operational version, $P_{a,b}(O)$, can be restricted to times up to the first measurement, yielding $P'_a(A)$; subsequently, the full applicable model can be reconstructed: $P_a(Q) = \sum_A P'_a(A) P_{a,f(A)}(Q|A)$.}

A more recent objection, that applies even to all-at-once accounts, has appeared in \textcite{wood2015}---although, notably, this stands as an objection to {\em all} accounts of entanglement phenomena, not specifically Type {\bf IIA} models.  The essential point is that causal channels are typically accompanied by signal channels, absent some special ``fine tuning'' of the underlying model.  Such fine-tuning would require additional explanation.  In any causal account of entanglement, such as the faster-than-light option of Type {\bf IA} models, {\bf signal locality} (the inability to send a spacelike signal) must be the result of some perfect cancellation in the marginal probabilities.  This is said to be ``fine-tuned'' because even a slight deviation would lead to spacelike signaling.  For example, in Quantum Field Theory, it is the perfect commutativity of spacelike-separated operators which guarantees the necessary ``fine-tuning.''

The situation might appear to lead to an additional challenge to Type {\bf II} models, with causal channels into the past, because another fine-tuning argument can be applied to {\bf signal causality} (the inability to send signals into the past).  But a more careful analysis reveals that the fine-tuning objection is not significantly worse for Type {\bf II} models than it is for Type {\bf I} models, because spacelike signaling violates {\bf signal causality} in some reference frame.  Further analysis of the Schulman model has revealed that the appearance of {\bf signal locality} follows from a basic symmetry \cite{almada2016}, providing just the sort of explanation (from symmetry) that is most often used to explain fine-tunings in high-energy physics.  A more comprehensive explanation of both {\bf signal locality} and {\bf signal causality} has also recently been proposed by \textcite{adlam2018a}.  Clearly, finding mathematical/physical principles underlying these signaling restrictions will be an important challenge for future reformulations of QM.

There is also a flip-side to the Wood-Spekkens fine-tuning argument.  If an underlying physics model indeed breaks time symmetry according to the {\bf NFID} condition, it would take a very finely balanced restriction to make microscopic physics look as time-symmetric as it does.  \textcite{leifer2017a} weigh this argument against the Wood-Spekkens fine-tuning argument, and propose that the time symmetry argument is stronger.

\subsection{Potential of Type {\bf IIA} models}
\label{sec:discussionII}

The examples of Section~\ref{sec:TypeIIA} demonstrate that a number of Type {\bf IIA} models can successfully account for the Bell-state correlations.  Thus, Bell's Theorem cannot be said to stand in the way of a locally-mediated reformulation of QM\@. In particular, the Schulman model admirably achieves a description in {\bf agreement} with QM which conforms to {\bf CA} ({\bf Continuous Action}), employing exclusively spacetime-based parameters with local, time-symmetric interconnections, which pose no difficulties for Lorentz covariance. 

A further dramatic advantage of such models relates to the exponential complexity of quantum states.  By using only spacetime-based parameters $Q$, the model $P_I(Q)$ has an evident physical interpretation: it specifies the probability of each \emph{possible} set of events in spacetime $Q$, while only one particular configuration actually occurs.  This is analogous to the Liouville equation in classical mechanics, where the statistical distributions can be exponentially complex, but only one phase-space configuration is taken to represent an actual physical system (even when we do not know which).  The complexity of this actual configuration scales linearly with the number of particles or the size of the modeled spacetime region.%
\footnote{Time plays a different role in the context of the Liouville equation, as within classical dynamics the configuration at one time determines the whole path.}
The Schulman model provides a simple example of such linear scaling, in that the parameters required for a two-particle experiment are merely two copies of the single-particle case. 

The exponential growth of the conventional wavefunction $\psi(t)$ with particle number might lead one to think 
that achieving such linear scaling would be impossible, especially if one viewed the information contained in $\psi(t)$ as some physical entity which had to be translated into parameters $Q(t)$ (the subset of $Q$ pertaining to a time $t$).  But note that $\psi(t)$ contains information about all possible measurement outcomes which might occur, for \emph{all possible} future measurement settings.  In an {\bf FID} model, $Q(t)$ can be a function of those future settings, and therefore need only inform the outcomes for the \emph{actual} future measurement, vastly reducing the required number of parameters [for further analysis, see \textcite{wharton2014}].

Beyond Bell state correlations, there are plenty of other quantum phenomena that must be addressed to approach a full reformulation of QM\@.  Single-particle interference appears challenging, but may be resolved in a Type {\bf IIA} model by adopting a field-based rather than a particle-based viewpoint \cite{wharton2018}.%
\footnote{Particle-like phenomena could arise from the discreteness of measurement interactions (the detector ``clicks''), enforced by boundary constraints, not by discreteness of the parameters.}
Recent Type {\bf IIA} models have tackled other issues, including position measurements of entangled particles \cite{sen2018}, and formal relativistic covariance \cite{wharton2010b,heaney2013,sutherland2017}.  Presumably more models will be developed in the near future, addressing additional issues such as 3-particle and partial entanglement phenomena.

There are many avenues which could be pursued in searches for such models.  Existing reformulations, such as Stochastic Mechanics \cite{nelson1966,nelson2012} and Stochastic Quantization \cite{damgaard1987}, could perhaps provide excellent starting points.  There are also some recent efforts which first evaluate the probabilities for the outcomes, $P_I(O)$ (using one of the standard methods of QM), and then define additional mediating parameters so that overall the resulting model is Type {\bf IIA} [both \textcite{sutherland2017} and \textcite{drummond2019} can be read in this manner].  While such approaches may claim applicability to a wide range of quantum phenomena, in our view, additional development is necessary for these models to fulfill their promise, such that the mediating parameters explain the outputs rather than the other way around.

There are quite a number of additional results in the literature which should guide the development of Type {\bf IIA} locally-mediated models.  Many of these have been developed in the context of locality [for a review, see \textcite{brunner2014}; a recent example is \textcite{carmi2018}].  A potentially important result which explicitly questions the arrow of time has recently been proven by \textcite{shrapnel2017}.  By dropping the usual {\bf NFID} assumption, their analysis indicates that such models must be ``contextual,'' meaning that distinct hidden-parameter accounts would be required for situations not distinguished by standard QM.  While it is not unreasonable to expect the details of intermediate hidden parameters to depend on the detailed intermediate context, it still raises the question of why standard QM cannot distinguish these differences.  This might indicate the development of models with inherent hidden symmetries, where this contextuality could seem more natural.

Eventually, Type {\bf IIA} models must also provide a satisfactory treatment of quantum measurements, but at the present stage of development this goal is not yet clearly in sight.  Still, the Type {\bf IIA} Schulman model improves upon standard Schr\"odinger-picture-with-collapse QM in two ways.  First, measurements do not correspond to any sudden collapse, so they look more like an ordinary interaction (the collapse-like event occurs somewhere between preparation and measurement).  Second, there is no confusion about whether the size of the relevant configuration space should expand (as in a QM interaction) or be reduced (as in a QM measurement), because nothing lives in configuration space; all parameters are associated with spacetime.

A future Type {\bf IIA} theory should provide an explanation for why the interaction between some large systems (measurement devices) and some smaller systems (such as the measured particles) can be described effectively by imposing boundary constraints on the smaller systems.  It is worth noting that such behavior is evident near large conductors in electromagnetism and thermal reservoirs in classical thermodynamics.  It is also well-known that smaller systems exhibit an evident time-symmetry in a way that larger, thermodynamic systems do not.  Understanding this is particularly important if time-symmetry is used as justification for introducing {\bf FID}, because this symmetry must somehow give way to the asymmetric {\bf signal causality} at larger, observable scales.

Taking the Schulman model as an illustrative example, the only time-asymmetry enters via a subtle distinction between photon preparations and photon measurements.  Both of these have controllable settings, but preparations have an additional point of control: the initial polarization is also treated as an input.  (For the case of entanglement, the initial correlation between two polarizations is an input.)  In contrast, the measurement does not allow this same level of control; one can input the final polarizer angle, but not the measured polarization (the latter is an output, not an input).  This empirically-based distinction between full control at preparations and mere setting control at measurements provides the symmetry-breaking mechanism which leads to the appearance of {\bf signal causality} in the model.  Everything else about the model respects time symmetry -- most notably, the intermediate account between preparation and measurement.

It is possible to attribute this distinction between preparation and measurement to the involvement of macroscopic ``agents,'' who have control of some quantities but not others \cite{price1997}.  Alternatively, one may attempt to include a description of the measurement process itself in the mathematical model.  Due to the observation that quantum measurements must have irreversibly recorded results [see, \emph{e.g.} \textcite{miller1996}], one should not expect a completely time-symmetric model to achieve this.  Future research into this issue may look in detail at the effects of a thermal environment, which could be included in a Type {\bf IIA} description.  In both classical and quantum cases, such treatments break time symmetry by fixing the initial states of the environment (averaging them over a known thermal distribution), while leaving its final states to be computed [see, \emph{e.g.}, \textcite{feynman1963}].  For an appropriate interaction between the system's degrees of freedom and the environment, the information regarding the values of some of the parameters pertaining to the ``measured'' system, $Q_M$, are effectively amplified and copied many times in the final state of the environment [see, \emph{e.g.}, \textcite{zurek2003}].  An intriguing possibility, called ``lenient causality'' in \textcite{argaman2018}, is that the time-symmetry-breaking in models of this type could impose {\bf signal causality} for parameters such as $Q_M$, without leading to {\bf NFID} for the microscopic parameters.%
\footnote{Better still, it could lead to a condition such as Information Causality \cite{pawlowski2009}, which is known to essentially guarantee compliance with the Tsirelson bound \cite{cirelson1980}.}

\section{Alternatives and Misconceptions}
\label{sec:alternatives}

While the previous two sections have discussed Type {\bf IIA} (locally-mediated) models in detail, there are many other models in the literature which can reproduce the experimentally observed CHSH-violations, including of course the existing formulations of QM\@.  This section will briefly discuss each of the general possibilities, giving references to some of the approaches not reviewed here.  These either violate {\bf Local Causality} in some different way, as categorized in Section~\ref{sec:categories}, or else they fall outside our framework, \emph{i.e.}, they are not in {\bf agreement} with QM as defined above.

Note that a specific approach can lead to a variety of models, and that a model must be fully specified in order to allow for a clear categorization.  (For example, as we have already seen in Section~\ref{sec:PoP} for the schematic models, a change in the spacetime location associated with $\lambda$ is sufficient to change the Type of the model.)  Below, we devote a subsection to each of the possibilities, and a final subsection to some misconceptions that might lead one to mistakenly believe there were additional categories of models.

\subsection{Type {\bf I} models}

Type {\bf I} models have no parameters that are dependent upon future inputs.  In Section~\ref{sec:aftermath} such models are categorized into Type {\bf IA} which would have faster-than-light mediators, and Type {\bf IB} in which distant regions can directly influence each other via non-spacetime-based (mathematical) intermediaries, such as the configuration-space wavefunction of conventional Schr\"odinger-picture QM\@.  The many-body wavefunction also enforces distant correlations in other Type {\bf IB} approaches, including Bohmian mechanics \cite{bohm1952} and spontaneous-collapse models \cite{GRW} (which only achieve full {\bf agreement} with QM in an appropriate limit).  Development of such models continues, \emph{e.g.}, with so-called ``flash''  models, which have parameters in spacetime (the flashes), but no intermediate screening parameters \cite{tumulka2006}.  

As noted in Section~\ref{sec:aftermath}, no representative Type {\bf IA} model has been formally developed [\textcite{norsen2010} might be the closest].  \textcite{spekkens2015} has noted that one can convert standard QM into a corresponding Type {\bf IA} model by introducing ``local copies'' of the wavefunction $|\psi(t)\rangle$ at every point in space with time coordinate $t$; ``collapse'' due to a distant measurement would then instantaneously update all of these new spacetime-based parameters.  Information is thus transferred from one region to another at an infinite speed, bypassing the $S''$ region of Figure 1(c), while passing through the upper boundary of the region $S'$ in Figure 1(b).

Whichever Type {\bf I} technique one uses to enforce correlations across spacelike separations, such a connection makes it difficult to achieve Lorentz covariance, even when {\bf signal locality} is satisfied.  In such models, when entanglement correlations between regions $\bm{1}$ and $\bm{2}$ are described, some observers see $\bm{1}$ affecting $\bm{2}$, while other observers see $\bm{2}$ affecting $\bm{1}$.  These descriptions do not properly transform into each other under Lorentz transformations,%
\footnote{The requirement here is not only that individual parameters transform covariantly, but that the overall description of which events affect which be consistent among different frames; see, \emph{e.g.}, \textcite{gisin2010}.}
motivating the possibility of omitting them altogether, resulting in a purely operational model, with just $P_I(O)$.  Despite these difficulties, it is clear that Type {\bf I} models are overwhelmingly represented in the relevant discussions in the literature.

\subsection{Type {\bf IIB} models}

While the previous sections have focused on Type {\bf IIA} models with spacetime-based mediators, other {\bf Future-Input Dependent} models can include non-spacetime-based entities, directly linking distant regions.  Such Type {\bf IIB} models often use configuration-space wavefunctions, in addition to their spacetime-based parameters.  Many of the above concerns about Type {\bf I} models (failure of Lorentz covariance, non-local influences, etc.) are therefore applicable to Type {\bf IIB} models as well. 

One popular Type {\bf IIB} model is the Two-State-Vector Formalism introduced by \textcite{aharonov1991}, which essentially doubles the state space of conventional QM.  For single-particle cases, it adds to the ordinary wavefunction $\psi(\bm{x},t)$ another wavefunction $\phi({\bm{x},t})$, a solution of the Schr\"odinger Equation which is determined by the setting and the outcome of the \emph{next} strong measurement on the particle (essentially a future boundary constraint).%
\footnote{Another similar example is the Transactional Interpretation \cite{cramer1980, cramer2016}, where the individual ``confirmation'' waves correspond to $\phi$.}  While these are naturally interpreted as spacetime-based parameters, for entanglement scenarios the relevant state vectors are conventional configuration-space wavefunctions, $\psi(\bm{x}_1,\bm{x}_2,t)$ and $\phi(\bm{x}_1,\bm{x}_2,t)$, and these entangled two-particle wavefunctions cannot easily be mapped onto spacetime-based fields. 
These wavefunctions are not spacetime-based, but are at least \emph{time}-based parameters, and in this generalized sense, they exhibit a violation of the essential ideas behind {\bf NFID}.  Having departed from spacetime, they no longer have any localized screening parameters, and so violate the {\bf CA} locality condition [see also \textcite{vaidman2013}].  It is therefore fair to categorize such a model as Type {\bf IIB}.

\subsection{Models outside the framework}
\label{sec:superdeterminism}

Various approaches in the literature raise more exotic possibilities, essentially claiming to not fall under any of the 4 model Types listed in Section~\ref{sec:categories}.  These approaches depart from our framework (Section~\ref{sec:framework}), either by violating the rules of probability theory, or by dropping aspects of the requirement of {\bf agreement} with QM\@.  The latter models risk losing the empirical content of QM, \emph{i.e.}, the comparison of $P_I(O)$ to experiment.  In order to still claim some form of agreement with QM, the $P_I(O)$ predictions must be recovered, at least at an effective level.  At that effective level, such models always fall within one of the Section~\ref{sec:categories} model Types.

One example is the Many Worlds Interpretation \cite{everett1957}, sometimes claimed as a way to avoid Bell's Theorem because all possible measurement outcomes are represented in a never-collapsed wavefunction.  In this approach, the measurement problem is avoided by removing the Born rule from the fundamental description, but then the empirical success of QM, the $P_I(O)$, is removed as well \cite{maudlin2010}.  Proponents of Many Worlds would argue that at an effective level, a version of the Born rule is still applicable, but the result is then a Type {\bf IB} effective model, in the same category as conventional QM\@. 

The deviation from our model framework which appears most frequently in the recent literature [perhaps because it was discussed repeatedly by \textcite{bell1981,bell1990,bell1977}] is ``superdeterminism,'' which retains the implicit {\bf NFID} assumption while considering violations of the ``{\bf $\bm{\lambda}$-independence}'' assumption, Eqn.~(\ref{eq:SI}).  This cannot be done within our framework (Section~\ref{sec:framework}), which treats the measurement settings $(a,b)$ as input parameters, corresponding to the mathematical concept of free variables.  But if the settings are treated as statistical parameters, {\bf $\bm{\lambda}$-independence} becomes
\begin{equation}
\label{eq:MeasInd}
 P_{c}(\lambda |a,b)=P_c(\lambda),
 \end{equation}
where $c$ encodes the free preparation setting, still treated as an input.  This is a statistical-independence relation, and permits a Bayesian inversion to an equation sometimes known as the ``no conspiracies'' assumption:
\begin{equation}
\label{eq:FreeWill}
 P_{c}(a,b|\lambda)=P_c(a,b).
 \end{equation}
Violations of this condition can then be pursued, by expanding $\lambda$ (or using additional variables) to include the systems that choose the measurement settings.

This approach has been seriously considered in the literature [\emph{e.g.}, \textcite{thooft2016}], despite the fact that it is only coherent if it makes sense to talk about the probabilities of the settings, $a$ and $b$.  But such probabilities cannot be defined without creating a conflict with standard QM, where $a$ and $b$ are free inputs.%
\footnote{Note also that the original suggestion, \textcite{shimony1976}, aimed only to emphasize the importance of the free-variable assumption, and argued that scientific exploration necessarily involves the assumption that ``hidden conspiracies of this sort do not occur.''  The reply of \textcite{bell1977} observed that even if the settings were chosen by a mechanical pseudorandom generator which could be included in an enlarged model, they would still be ``effectively free for the purpose at hand.''}
Indeed, in the explicit superdeterministic toy-models which have been proposed for the Bell-state correlations, the relevant hidden variables [$\lambda_0$ in \textcite{brans1988}, and $\mu$ in \textcite{hall2016}] are simply copies of the measurement setting parameters $a$ and $b$, transferred to earlier times.  The other elements of these models prescribe a specific form of $P_{a,b}(\lambda)$ and a role for $\lambda$ in generating the outputs, as discussed in Section~\ref{sec:Hall}.  In practice, therefore, explicit superdeterminstic models which agree with QM are forced to treat the future settings $a$ and $b$ as free inputs.  Once this is acknowledged, the model again falls within the framework, and its Type can be identified.

There are additional well-established methods which can be more spacetime oriented, but do not meet the probability rules of our framework.  For example, path-integral accounts of QM utilize spacetime-localized paths.  It might be tempting to think that each path might be represented by a set of parameters $Q$, but the path integral cannot be parsed into normalized probabilities $P_I(Q)$ where only one path $Q$ can be taken to exist.\footnote{Introducing an {\bf FID} viewpoint, along with a different parsing of $Q$, might potentially resolve this problem \cite{wharton2016}.}  Similarly, Quantum Field Theory (QFT) can be viewed as assigning a complex amplitude to all possible field configurations in spacetime, but each of these configurations cannot be assigned a probability.  A further example is given by the consistent histories approach [\emph{e.g.}, \textcite{griffiths2011}], where the probability rules for the intermediate description $P_I(Q)$ are changed, while those for the outputs, $P_I(O)$, are not.  These approaches represent directions which are, in a sense, more radical than the search for Type {\bf IIA} models.

\subsection{Misconceptions}
\label{sec:misconceptions}

It has often been claimed that Bell's Theorem is based on additional assumptions not identified above, including \emph{determinism} and \emph{realism}.%
\footnote{Once the discussion is cast purely in terms of mathematical models as done here, assumptions of ``realism'' can play no role [see \textcite{norsen2007} for a discussion in a wider context].  Note that when ``realism'' is taken to imply that systems have properties prior to measurements, the {\bf NFID} assumption is again being taken for granted, assuming not only that the systems have ``objective'' properties, but also that these properties are independent of the settings of future measurements.}
These erroneous claims are already well-addressed in the literature \cite{norsen2007, norsen2011, norsen2017, maudlin2010, maudlin2014}, but some clarifications will be repeated here in order to alert the reader to some of the many controversies in the literature.

Bell did not originally present his proof as outlined in Section~\ref{sec:theorem}; this unified approach only came later.  The EPR paper (see Section~\ref{sec:interlude}), had already demonstrated that certain perfect correlations between distant measurements clearly violate {\bf Local Causality}, unless one adds deterministic hidden parameters.  \textcite{bell1964} built upon this result, and showed that even with deterministic hidden parameters {\bf Local Causality} could not be saved, as other predictions of QM could not be obtained.

Unfortunately, the argumentation of EPR contained several additional elements which made it appear paradoxical even before Bell's work, and the notion that \textcite{bohr1935} had refuted it was widespread [see, \emph{e.g.}, \textcite{clauser1969}].  As \textcite{bell1964} did not
go through the EPR part of the argument in any detail \cite{wiseman2014,norsen2015}, many have concluded that the implications could be avoided by not postulating hidden parameters in the first place, or by not requiring them to be ``deterministic'' (or ``realistic'', or ``counterfactual definite'', etc.).  But such moves do not save {\bf Local Causality}, for the reasons given in the EPR paper.  Bell himself later wrote: ``It is remarkably difficult to get this point across, that determinism is not a \emph{presupposition} of the analysis'' (Bell, 1981; emphasis in original)\nocite{bell1981}.%
\footnote{
The original derivation of the CHSH inequality \cite{clauser1969} simply assumed deterministic hidden parameters, without using the EPR argument.  It was rapidly understood that the same inequality also holds for indeterministic local hidden-variable models [see footnote 10 of \textcite{bell1971}, \textcite{clauser1974}, or the unified type of proof as in Section~\ref{sec:CHSH}], but this is often ignored.}
It is hoped that the explicit discussion of the framework and assumptions in the present work will help alleviate such difficulties.

\section{Conclusions}

This Colloquium began by carefully framing the assumptions that lead to Bell's Theorem, in terms of input-parameters $I$ and non-input parameters $Q$, both associated with locations in space and time.  By defining a model in terms of the probabilities $P_I(Q)$ which it generates, Bell's Theorem indicates that any such model which is in agreement with QM must violate one of the original assumptions, one of the components of {\bf Local Causality}.  This allows a natural categorization of all possible reformulations of QM, as described in Section~\ref{sec:categories}.  

To the extent we require the parameters in our mathematical models to correspond to \emph{physical events}, this {\bf Local Causality} violation is quite significant.  Einstein described the physical justification for Local Causality in a 1948 letter \cite{born1971}:
\begin{quote}
If one asks what, irrespective of quantum mechanics, is characteristic of the world of ideas of physics, one is first of all struck by the following: the concepts of physics relate to a real outside world...  It is further characteristic of these physical objects that they are thought of as arranged in a space-time continuum. An essential aspect of this arrangement of things in physics is that they lay claim, at a certain time, to an existence independent of one another, provided these objects `are situated in different parts of space'.

The following idea characterizes the relative independence of objects far apart in space (A and B): external influence on A has no direct influence on B ...
\end{quote}

But Bell showed that this line of thinking leads to limitations on distant correlations which are in direct conflict with QM\@.  The outcomes of spatially-separated experiments are correlated in a manner which cannot be explained only in terms of common past inputs.  Still, it does not follow that our only option is to throw out the entirety of Einstein's analysis, giving up on ``physical objects... arranged in a space-time continuum''.  At least one of the assumptions that make up {\bf Local Causality} needs to go, but spacetime-associated parameters might still be retained.  Indeed, if they are not retained to some extent, all concepts of ``locality'' lose their usual meaning.

One concept of locality in particular, {\bf Continuous Action}, is defined above in a time-neutral manner that prevents unmediated ``action-at-a-distance''.  Even given Bell's Theorem, this definition of locality can be retained in two different styles of quantum models, categorized as Type {\bf IA} and Type {\bf IIA}.%
\footnote{The different Types are identified in Table I of Section III.A.}
The former would require faster-than-light mediating parameters, so only the latter is compatible with Lorentz covariance.  The price for retaining Lorentz covariance while forbidding action-at-a-distance is the violation of an assumption arguably unrelated to locality: the premise that a model's parameters should not functionally depend on inputs associated with the future of those parameters, or {\bf No Future-Input Dependence}.  Without this assumption [or its corollary, Eqn.~(\ref{eq:SI}), {\bf $\bm{\lambda}$-independence}] Bell's Theorem cannot be derived.  

This analysis therefore motivates Type {\bf IIA} models with {\bf Future-Input Dependence} and {\bf Continuous Action} as the most ``local'' models compatible with QM.  Very roughly, these models would violate {\bf Local Causality} by violating our intuition of ``causality'' rather than our intuition of ``locality''.\footnote{See Section~III.B for clarification of these issues.}  
Einstein saw no reason to relax either one of these, and Bohr effectively relaxed both, taking an operational view which keeps only the {\bf signal causality} and {\bf signal locality} conditions.  Bell and his followers took the ``causality'' condition for granted, without realizing that an alternative exists,%
\footnote{When prompted to consider the failure of {\bf $\bm{\lambda}$-independence}, Eqn.~(\ref{eq:SI}), which they called measurement-independence, they always considered the conspiratorial superdeterministic option, discussed in section VI.C.}
and as a result studied Type {\bf IA} and Type {\bf IB} models.  Others took an operational approach which drops both the ``causality'' and the ``locality'' requirements for the internal (hidden) variables, resulting in development of Type {\bf IIB} models.

As analyzed in Section IV, Type {\bf IIA} models of quantum entanglement have effective connections associated only with the particle world-lines, either within the lightcones or on the lightcones for photons (\emph{i.e.}, there are no direct space-like connections).  Dropping the {\bf No Future-Input Dependence} assumption allows these to be effective \emph{two-way} connections.  Using this strategy, Einstein's ``independence of objects far apart in space'' can be softened without requiring connections which violate the spirit of relativity.  In particular, this view accommodates entanglement scenarios by allowing an external influence on A to have an \emph{indirect} influence on B, via mediating events in the intersection of their past lightcones, without raising any difficulties with Lorentz covariance.  As discussed in Section~\ref{sec:con_arguments}, this need not lead to logical inconsistencies, or deviations from conventional QM predictions.

Physics models with explicit {\bf Future-Input Dependence} have been developed already in the context of classical Electrodynamics \cite{wheeler1945,wheeler1949}, and their relevance to Bell-like scenarios was pointed out even before Bell's Theorem emerged \cite{costa1953}, and then repeatedly since [\emph{e.g.}, \textcite{pegg1982,price1997}].  Despite this, the development of explicit Type {\bf IIA} models of entanglement phenomena has only recently begun in earnest, and is currently limited to a few particular applications, most notably the Bell state correlations which typically serve to demonstrate the issue of Bell's Theorem.  The detailed discussion of the proof-of-principle examples of such models in Section~\ref{sec:TypeIIA} is hoped to introduce these possibilities to a wider audience, and Section~\ref{sec:discussionII} indicates several possible avenues for future developments.  This would include describing more complicated entanglement scenarios and developing a treatment of quantum measurements as interactions between small and large systems. 

It is emphasized that while {\bf Future-Input Dependent} (or ``retrocausal'') models of QM can have an underlying structure that is as time-symmetric as classical physics, all such models must have a mechanism to recover the time-asymmetric condition of {\bf signal causality}.  Two possibilities for such a mechanism have been suggested above.  The first emphasizes the role of time-asymmetric ``agents'' employing the theory: they select which parameters of a theory to use as inputs of a specific model and which as outputs.\footnote{See \textcite{price1997} for further discussion.}  The second considers the possibility of a time-symmetry-breaking physical principle (perhaps due to the low entropy of the big bang), with possibly relatively minor effects on the mathemtical model, \emph{e.g.}, a specification of initial conditions.  As a result of this mild symmetry breaking, irreversibility could appear in the thermodynamic limit, and with it, {\bf signal causality}.\footnote{It is interesting to note that \textcite{bell1990} already asked: ``Could it be that causal structure emerges only in something like a `thermodynamic' approximation?''  But his tentative answer was negative, possibly due to his taking {\bf NFID} for granted.}

A successful Type {\bf IIA} reformulation of QM would employ only spacetime-based parameters and would associate conventional probabilities with each fully-specified configuration.  An appropriate interpretation would take only one of these possibilities to actually occur in Nature.  In other words, the number of parameters describing a system would grow only linearly with its extent.  This stands as a dramatic advantage over existing approaches, where the number of necessary parameters scales exponentially with the number of particles in the system.  Combined with {Lorentz covariance}, this could greatly alleviate the disconnect between quantum theory and general relativity.

Such a reformulation would also shed light on an unresolved issue in quantum foundations---how to interpret the conventional wavefunction $\psi$ and the collapse postulate.  Although $\psi$ is not included in the underlying model, it could still represent available \emph{knowledge} about the actual parameters---a viewpoint that has become known as ``$\psi$-epistemic'' \cite{spekkens2007}.  Such states of incomplete knowledge naturally reside in configuration space (as in classical statistical mechanics), as they have to represent a large number of possible correlations.  Unitary evolution of these states would then correspond to time-evolving the available information, in analogy to Liouville dynamics.  Learning additional information about future settings and future outcomes would then lead to a Bayesian updating of $\psi$, corresponding to a (non-physical) collapse.   This is essentially the style of model advocated by Einstein, where the actual state of the system was not $\psi$, but rather something more fundamental \cite{harrigan2010}.  

While the present work is focused on Bell's Theorem, additional lines of research are also converging on the promise of {\bf Future-Input Dependent} models.  As discussed above, \textcite{leifer2017a} motivate such models via time symmetry.  Another argument is motivated by the much-discussed Pusey-Barrett-Rudolph (PBR) Theorem \cite{pusey2012}, recently reviewed by \textcite{leifer2014}, and yet another relies on arguments concerning the complexity achievable with quantum computation \cite{argaman2019}. One of Leifer's conclusions exactly matches ours, promoting the development of ``retrocausal $\dots$ models that posit a deeper reality underlying quantum theory that does not include the quantum state.''  The spacetime-associated parameters $Q$ in {\bf Future-Input Dependent} models would mathematically represent this ``deeper reality''.  Fully realizing this goal remains an open challenge.

{\bf Acknowledgements:} The authors warmly thank A.~Briggs, P.~Drummond, J.~Finkelstein, S.~Friederich, R.~Sutherland, and several anonymous Referees for useful discussions and comments on a draft of the manuscript. This work is supported in part by the Fetzer Franklin Fund of the John E. Fetzer Memorial Trust.

\section*{Appendix: Derivation of the Schulman Model}

Schulman's original single-particle model applies to a single spin-$1/2$ particle; here we convert it to a photon polarization problem.  The photon's classical trajectory is known, and it has a real (hidden) polarization direction $q(t)$ everywhere on its trajectory.  The photon is prepared and measured {by} passing through two polarization cubes, with the first set at an angle $\theta_1$ and the second set at $\theta_2$.  The initial polarization is constrained, $q(t_1)=\theta_1$, as is usual for initial boundary conditions.  Schulman enforced a similar final boundary condition at measurement, where the final polarization was constrained to be either $q(t_2)=\theta_2$ or $q(t_2)=\theta_2+\pi/2$.  

This final constraint is controllable (modulo $\pi/2$) and the model is FID.  The time-asymmetry (modulo $\pi/2$ at the output, but modulo $\pi$ at the input) is external: an experimenter can choose to block a photon with an unwanted input polarization, but does not know the output polarization until it is too late to interfere.  Otherwise, everything in this model is fully time-symmetric. 

Such two-time-boundary problems can only be solved ``all-at-once,'' with probabilities assigned to entire histories, $q(t)$, not instantaneous states.  (One can extract the latter probabilities from the former.)  Defining a net rotation 
\begin{equation}
\label{eq:Dl}
\Delta q\equiv\int^{t_2}_{t_1} \frac { d q(t)}{dt} dt
\end{equation}
(which is permitted to be larger than $2\pi$ for multiple rotations), the convolution of Schulman's proposed Cauchy kicks imply the probability assignment of Eqn.~(\ref{eq:SProb}):
\begin{equation}
\label{eq:SProbA}
P(\Delta q)\propto\frac{1}{(\Delta q)^2+\gamma^2}.
\end{equation}
Remarkably, this distribution recovers Malus' Law as $\gamma\to0$. Seeing this requires adding the probabilities for all the rotations which end at the same polarization angle (modulo $\pi$), and normalization.

The evaluation requires summing over all the possibilities of getting from $\theta_1$ to $\theta_2$(mod $\pi$), allowing for rotations through angles larger than $\pi$ in both directions.  The sum, 
$\sum _{n=-\infty }^{\infty } \frac{1}{( \Delta\theta + n\,\pi )^{2}}$
with $\Delta\theta = \theta_1 - \theta_2$,
can be calculated, similarly to Euler's solution of the Basel problem ($\sum _{n=1 }^{\infty } \frac{1}{n^{2}}$), by equating two different families of polynomial approximations to the same function, in this case, 
$f(x)=\sin(\Delta\theta+x) \sin(\Delta\theta-x)$.  
One family is the Taylor expansion, and as 
$f(x)=\frac{1}{2} \left( \cos(2x) - \cos(2 \, \Delta\theta) \right)$, the coefficient of $x^2$ is $-1$, yielding $f(x) = \sin^2(\Delta\theta) - x^2 +O(x^4)$.

The other polynomial approximation scheme is obtained by multiplying the value of the function at $x=0$ by a factor of $(1-x/z_k)$ for each of the zeros $z_k$ (roots) of the original function (a specific approximation is obtained by including only roots up to a certain absolute magnitude).  Treating the roots in pairs, $z_n=-z'_n=\Delta\theta+n\pi$, gives 
$f(x) = \sin^2(\Delta\theta) \, 
\Pi_{n=-\infty}^\infty \left( 1 - \frac{x^2}{(\Delta\theta+ n\pi)^2} \right)$, 
and expanding only up to terms quadratic in $x$ gives the necessary sum:
\begin{equation}
\label{eq:Sch_sum}
\sum _{n=-\infty }^{\infty } \frac{1}{( \Delta\theta + n\,\pi )^{2}} = 
\frac{1}{ \sin^{2} \left( \Delta\theta \right) } .
\end{equation}
Normalizing the probabilities for either $q(t_2)=\theta_2$ or $q(t_2)=\theta_2+\pi/2$, is achieved by simply multiplying by the product of the corresponding denominators on the right-hand side of \eq{eq:Sch_sum}, yielding Malus' Law: $p=\cos^2(\Delta\theta)$, as required for Section~\ref{sec:Schulman}.

Remarkably, Schulman also used this idea to \emph{prove} the Born rule, in the sense of showing that probabilities $\propto |\psi|^x$ are compatible with this idea of multiple kicks only for $x=2$ (whether or not the Cauchy-Lorentz distribution is used for each kick).

\bibliography{RMPBibTexFile}

\end{document}